\documentclass[%
british,
reprint,
superscriptaddress,
 amsmath,amssymb,
 aps,
]{revtex4-2}
\usepackage{graphicx,color,subfigure}
\usepackage{amssymb,amsmath} 

\begin{document}

\title{Stiffening attractive suspensions with solid inclusions or granularity effect in loaded colloidal gels}

\author{Claudia Ferreiro-Córdova}
\affiliation{Université Paris-Saclay, CNRS, Laboratoire de Physique des Solides, 91405, Orsay, France}
\affiliation{Tecnologico de Monterrey, Escuela de Ingeniería y Ciencias, Querétaro, Querétaro, 76130, Mexico}

\author{Giuseppe Foffi}
\affiliation{Université Paris-Saclay, CNRS, Laboratoire de Physique des Solides, 91405, Orsay, France}

\author{Olivier Pitois}
\affiliation{Université Gustave Eiffel, Ecole des Ponts ParisTech, CNRS, Laboratoire Navier, F-77447 Marne-la-Vallée, France}

\author{Chiara Guidolin}
\affiliation{Université Paris-Saclay, CNRS, Laboratoire de Physique des Solides, 91405, Orsay, France}

\author{Maxime Schneider}
\affiliation{Université Paris-Saclay, CNRS, Laboratoire de Physique des Solides, 91405, Orsay, France}

\author{Anniina Salonen}
\affiliation{Université Paris-Saclay, CNRS, Laboratoire de Physique des Solides, 91405, Orsay, France}

\date{\today}

\begin{abstract}

The elastic properties of a soft matter material can be greatly altered by the presence of solid inclusions whose microscopic properties, such as their size and interactions, can have a dramatic effect. In order to shed light on these effects we use extensive rheology computer simulations to investigate colloidal gels  with solid inclusions of different size. We show that the elastic properties vary in a highly  non trivial way as consequence of the interactions between the gel backbone and the inclusions. In particular, we show that the key aspects are the presence of the gel backbone and its mechanical alteration originated by the inclusions. To confirm our observations and their generality, we performed experiments on an emulsion that presents strong analogies with colloidal gels and confirms the trends observed in the simulations.
\end{abstract}

\maketitle

\section{Introduction}

Composite materials have been created since antiquity to combine and enhance the mechanical properties of the constituents. Since Einstein thought about the effects of a low concentration of solid particles on the viscosity of Newtonian fluids, experimental and theoretical works have explored the mechanical properties of a wide range of such materials.

The types of matrix and particle that can be used are enormous and the variability in the properties of the resulting materials equally. For example, the addition of nanoparticles or fibrous objects can be highly efficient in changing the elastic modulus, because of the high surface area of nanoparticles and the capacity  of the fibres to create percolated networks even at low volume fractions\cite{composite_nanos}. In the absence of specific particle-matrix interactions, classical mechanics can be used to estimate the elastic modulus of such composite materials. Those are based on the assumption that the matrix is a continuous linear elastic medium embedding ellipsoidal inclusions \cite{Eshelby_1957, Eshelby_1959}. The Krieger and Dougherty (KD) equation \cite{KriegerDougherty_1959} was found to describe well the elastic modulus of various elastic matrices loaded with monodisperse or bimodal non-Brownian spherical particles \cite{Mahaut2008,Vu_2010}, although the hypothesis of continuous material around the inclusions was not fully justified for some of the studied matrices such as emulsion or bentonite. This could be understood by the fact that the inclusions were large compared to the characteristic length-scale of those matrices. Recent work with liquid foam loaded with solid particles showed that the elastic modulus was correctly predicted by the KD equation for inclusion-to-bubble size ratios close to unity \cite{Gorlier_SoftMatter_2017}. On the other hand, it has been shown that if the particles are small enough to concentrate between the bubbles and to form a skeleton-like granular packing, stronger strengthening can be obtained \cite{Gorlier_PRE_2017,Addad_PRL_2007}. When the particles were attractive, and formed a gel within the foam channels, bulk gel elasticity was not sufficient to describe the strengthening as the gel. This is because the gel time (time at which an elasticity can be measured) changes with confinement \cite{Foam_gel_Alesya}. 

Those results revealed that finite size effects are not easy to anticipate for a particle loaded matrix, especially if the matrix can form various different micro-structures. This is the case for colloidal gels for which the individual particles aggregate and cluster to form stress-bearing networks of gels. At low volume fractions ($\phi_g < 0.1$) it is understood that the particles form clusters which then percolate \cite{Weitz_PRL_1984}. At high volume fractions ($\phi_g > 0.5$) the arrested states are spatially homogeneous, the particles are localised between their neighbours and can be described as attractive glasses \cite{Pham2002}. Those structures can be described using the cluster size as the characteristic length scale at low $\phi_g$, whereas the particle size is the relevant size at high $\phi_g$. The transition from one to the other was predicted by Zaccone et al \cite{Zaccone_PRL2009}, where they also showed how the cluster-cluster interactions control gel elasticity. The role of the number density of the clusters connections was further explored by Whitaker et al \cite{Whitaker_NatComm_2019}.

The aim of this work is to discuss how colloidal gels' mechanical properties are modified by the presence of non interacting spherical inclusions of different sizes. In the first part of the manuscript, we will present a simple numerical model in which the colloidal gel is the result of an arrested phase separation~\cite{foffi2005, Zaccarelli2007}. Using a finite strain rate algorithm~\cite{colombo2014, FerreiroCrdova2020}, we measured the mechanical properties as a function of volume fraction of the inclusions as well as of their size. The picture that emerges suggests that the size ratio between the particle of the gel and the inclusions plays a major role. In fact, when the inclusion's size is comparable to the size of the gel particles, the resulting composite gel is noticeably reinforced and can sustain much larger stresses. We will show that this effect results from a non trivial coupling between the structure and local mechanical stress that results in an imbalanced distribution of local stresses. In a second part of the paper, we will present an experimental model of a colloidal gel, made of emulsion drops, reinforced with solid polystyrene particles and we show a semi-quantitative agreement with the simulation predictions.  The similarity between the results from the experimental and numerical models, despite their intrinsic differences,  suggests that the reinforcing effect that we observe in our composite gel could be general and translated in other similar systems.

\section{Methods}
\label{sec1}

In this paragraph we will introduce the numerical approach, as well as the experimental system.

\begin{figure}%[]
\centering
\includegraphics[width=0.3\textwidth]{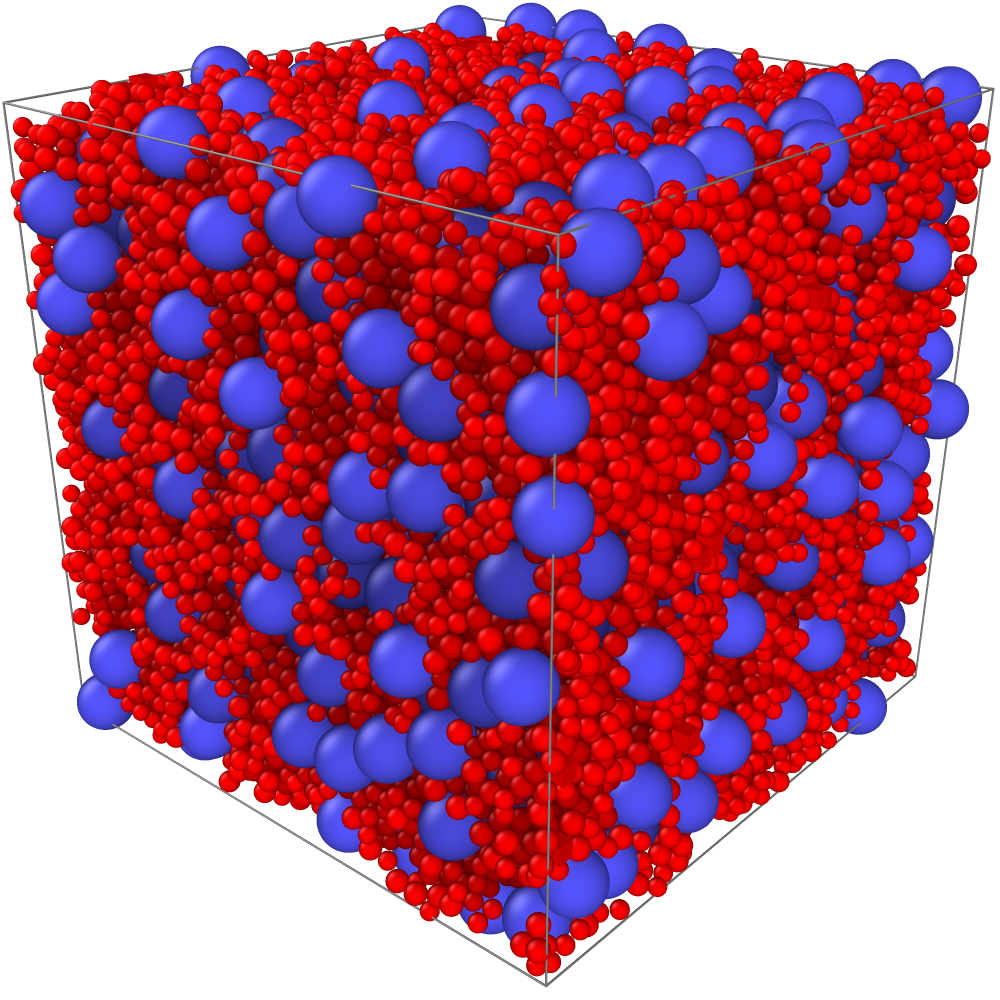}
 \caption{Snapshot showing a typical simulation box for a mixture with a ratio $\sigma \!=\!3$ for the inclusion particles. The inclusion and gel particles are at equal relative concentration, with $\phi_{T}\!=\!0.60$, $ \phi_{g}\!=\!0.30$ and $\phi_{inc}\!=\!0.30 $. }
\label{fig:Images}
\end{figure}

\subsection{Simulations}

\subsubsection{Numerical model}

In our simulations, we use a  colloidal  mixture model made of $N$ spherical particles with two distinguishable species: gel particles and inclusion particles. The particles tagged as \textit{g} represent the building block of the gel backbone, which is formed by particles with two different diameters, at equal relative concentrations, that have an average diameter of $\langle D_{g} \rangle$. Particles tagged as \textit{inc} model the spherical solid inclusions of diameter $D_{\text inc}$. The interactions are described by the potential

\begin{equation} \label{eq:potentiallq}
u_{s(i) s'(j)}(r_{ij})\!=\!\! \left\{
  \begin{array}{l l}
     \!\!C \epsilon \! \left[\! \left( \!\!\frac{D_{ij}}{r_{ij}}\!\! \right)^{p}\!\!-\!\left( \!\!\frac{D_{ij}}{r_{ij}}\!\! \right)^{q} \right]\!\!+\! U_{ss'}\!\!\!\! & \quad r_{ij}\!\! <\!\! r^{ss'}\\\\
     \!\!0 \!\! & \quad r_{ij}\!\! \geq\!\! r^{ss'}
  \end{array}\!\!, \right.
\end{equation}
where $r_{ij} \!~=~\! |\mathbf{r_{j}} - \mathbf{r_{i}}|$ denotes the distance between two particles ($i$ and $j$), $D_{ij}$ is the average diameter of the interacting particles $i$ and $j$, $s(i)$ is the species of particle $i$ and $s'(j)$ is the species of particle $j$. The constant $U_{ss'}$ is chosen accordingly such that $u_{ss'}(r^{ss'})=0$. The exponents $p$ and $q$ in Eq.~\ref{eq:potentiallq} define the range of the attraction, and are set to $p\!~=~\!36$ and $q\!~=~\!24$. The constant $C\!~=~\!\frac{p}{p-q}\left(\frac{p}{q}\right)^{q/(p-q)}$ ensures that the minimum of the potential is unitary while the cut-off parameter $r^{ss'}$ sets the range of the interaction. For simplicity, we have set $\epsilon\!~=~\!1$.

The cut-off $r^{s s'}$ is chosen such that
\begin{equation} \label{eq:potentialcut}
r^{s s'}= \left\{
  \begin{array}{l l l}
    1 .9 \, D_{ij} &  s\!=\!s'\!=\!1& {\rm gel-gel}\\\\
%     1.03d_{ij}& {\rm else} & 
     D_{min}& {\rm else} & 
  \end{array}, \right.
\end{equation}
where $D_{min}=(3/2)^{1/2} \, D_{ij}$ is the distance at which the potential in Eq.~\ref{eq:potentiallq} has its minimum. Then, with Eq.~\ref{eq:potentialcut}, the potential in Eq.~\ref{eq:potentiallq} describes gel-gel interactions as attractive while all other interactions as completely repulsive.

We have chosen the two different diameters of the gel particles as $D_{g1}=0.909\, D_0$ and $D_{g2}=1.091\, D_0$. With this, the average diameter of the $N_g$ gel particles is $\langle D_{g} \rangle=(1/2)( D_{g1} +D_{g2}) = D_{0}$. The unit of lengths are fixed as  we have set $D_0=1$. To characterize the size of the inclusions, we use the ratio $\sigma$ between the diameter of the inclusion particles and the average diameter of the gel particles, $\sigma=D_{inc}/\langle D_{g} \rangle$. The ratio values $\sigma$ studied here are 0.8, 1.0, 1.3, 2.0, 3.0, 4.0, 5.0 and 10.0. The total packing fraction of our simulation box is set as $\phi_T=\phi_{g} + \phi_{inc}$, with $\phi_{g}$ and $\phi_{inc}$ the effective packing fraction of the gel and inclusion particles in the simulation box. To study the effect of different loads of inclusion particles in the final composite structures, we chose the total packing fraction range $0.30 \leq \phi_T \leq 0.60$, with $0.10 \leq \phi_{g} \leq 0.50$ and $0.10 \leq \phi_{inc} \leq 0.40$. The total number of particles in our systems is $1.4 \times 10^{4} \leq N \leq 6.6 \times 10^{4}$.

All simulations were performed with the open-source MD simulation package LAMMPS\cite{plimpton1995}. For each total packing fraction value studied, we start with an initial isotropic configuration, with no overlaps, which we generate randomly at the desired total packing fraction. After this, the interaction described by Eq.~\ref{eq:potentiallq} is turned on and the system is allowed to evolve at $k_BT/\epsilon \!~=~\! 0.01$ for a period of at least $4\times 10^{3} \, \tau^*$, where $\tau^*\!~=~\!\sqrt{m(D_0)^2/\epsilon}$ is the molecular dynamics time unit in our simulations. The time step in the initial sample preparation runs is $\delta t \!~=~\!5 \times 10^{-4}\tau^*$. After the initial runs, we apply the following dissipative viscous dynamics to our system\cite{colombo2014}
\begin{equation}\label{eq:damp_dynamics}
m \frac{\mathrm{d^2\mathbf{r}_i}}{\mathrm{dt^2}}=-\xi \frac{\mathrm{d\mathbf{r}_i}}{\mathrm{dt}}-\nabla_{\mathbf{r}_i} u.
\end{equation}
Where $m$ is the particle mass and $\xi$ the friction coefficient. We use $m/\xi\!~=~\! 1.0 \tau^*$, which corresponds to the overdamped limit~\cite{Nicolas2016}. This dynamics is run for at least $1\times 10^{4} \, \tau^*$, which draws the kinetic energy  to a value lower than $10^{-10}$ and leads to an arrested configuration. From this stage onward, all time steps of the simulations are set to $\delta t \!~=~\!5\times 10^{-3} \, \tau^*$.

In Fig~\ref{fig:Images}, we  present a typical final arrested configuration for a gel network (red particles) with solid inclusions (blue particles) at a total packing fraction $\phi_{T}=0.60$ and equal relative concentrations, $\phi_{g}\!=\!0.30$ and $\phi_{inc}\!=\!0.30$, with a ratio $\sigma=3$ for the inclusions. Here it can be seen that the solid inclusions are scattered in the simulation box, without any apparent preference, and that the gel particles are forming a network. In the following sections, we will expand more on the composite structures explored.

\subsubsection{Mechanical tests} 

The mechanical properties of our composite systems were tested using continuous strain deformation and small amplitude oscillatory rheology (SAOR), as previously reported for colloidal gels~\cite{FerreiroCrdova2020}. To study samples under continuous strain deformation, we applied a series of incremental strain steps, in simple shear geometry, to each sample analysed\cite{colombo2014,FerreiroCrdova2020}. This is done by first applying, for each step, the instantaneous affine deformation $\Gamma_{\Delta \gamma}$ to the simulation box,
\begin{equation}\label{eq:def_strain}
\mathrm{\mathbf{r}}'_{i}=\Gamma_{\Delta \gamma} \mathrm{\mathbf{r}}_{i}=\begin{pmatrix}
 1 & \Delta \gamma & 0 \\
 0 & 1 & 0 \\
 0 & 0 & 1
\end{pmatrix}\mathrm{\mathbf{r}}_{i}.
\end{equation}
which represents  a simple shear in the $xy$ plane. After a strain step deformation of $\Delta \gamma$ is imposed, the boundary conditions are updated and the simulation box is then relaxed by applying the damped dynamics described in Eq.\ref{eq:damp_dynamics}. This procedure is repeated for $n$ steps, obtaining a cumulative strain in the system of $\gamma=n\Delta\gamma$. We have chosen a strain increment of $\Delta \gamma\!~=~\!0.001$ and a relaxation time interval $\Delta t$ such that the shear rate is $\dot{\gamma}=\Delta \gamma/ \Delta t \!~=~\!10^{-5}\tau_0^{-1}$, with $\tau_0=\xi D_0^2/\epsilon$ the time it takes a particle subjected to a force $\epsilon/D_0$ move a distance of $D_0$.

The average state of stress of the system is given by the tensor $\Sigma$. In our simulations we compute each component $\Sigma_{\alpha\beta}$ using the standard viral definition\cite{colombo2014,FerreiroCrdova2020},
\begin{equation}
\Sigma_{\alpha\beta}=\frac{1}{V} \sum_{i=1}^{N} \frac{\partial u}{\partial r_{i}^{\alpha}} r_{i}^{\beta}.
\end{equation}
where $\alpha,\beta$ stand for the Cartesian components $\{x,y,z\}$. In this analysis, we consider only the pairwise energy contribution, all other terms are assumed negligible as the velocities for all particles are kept small for all shear rates. For the step-strain curves, we compute the shear component $\Sigma_{xy}$ which corresponds to the same component of the deformation. In the following sections we refer to $\Sigma_{xy}$ as $\Sigma$. 

For our SAOR analysis, the frequency dependent response is measured by imposing the oscillatory shear strain  $\gamma(t)\!~=~\!\gamma_0\sin(\omega t)$ to the system, with the dynamics given by
\begin{equation}
m \frac{\mathrm{d^2\mathbf{r}_i}}{\mathrm{dt^2}}=-\nabla_{\mathbf{r}_i} u- \xi \left( \frac{\mathrm{d\mathbf{r}_i}}{\mathrm{dt}}- \dot{\gamma}(t) y_i \mathbf{e_x} \right).
\end{equation}
During each cycle, we compute the shear stress component $\Sigma$, as for the step-strain analysis. The first harmonic expansion of the shear stress, for a visco-elastic solid, can be written as:
\begin{align}\label{eq:liss_harmonics}
\Sigma(t) =& \gamma_0 \Big[ G^{\prime}(\omega, \gamma_0) \sin (\omega t)  + G^{\prime \prime}(\omega, \gamma_0) \cos (\omega t) \nonumber \Big].
\end{align}
With $G'(\omega, \gamma_0)$ and $G''(\omega, \gamma_0)$ the elastic and plastic moduli, respectively. In the present work we will focus our analysis on $G'$, which is refered to as $G$ in the following sections. The modulus is extracted by a simple Fourier transform of $\Sigma$ and $\gamma$.

\subsection{Experiments}
\label{Sec: Meth_Exp}

We have made weakly elastic emulsions of rapeseed oil (from Brassica Rapa, Sigma-Aldrich) in aqueous solutions of sodium dodecyl sulphate (SDS, Sigma-Aldrich) at 30 g/L. The surfactant solution is freshly prepared by dissolving SDS in deionised water and used within one day. The emulsions are made using the double syringe method \cite{Gaillard_generation}: two syringes (60 mL, Codan Medical) are partially filled with oil and SDS solution respectively and connected with a double luer lock. The oil and the aqueous phase are then manually pushed through the connection 20 times, at which point a homogeneous emulsion is formed.
A stock emulsion at 0.70 of oil is prepared with an average drop diameter, $2\langle R_g \rangle$ of 5 $\mu$m. Emulsions at lower oil volume fractions, between 0.63 and 0.44, are obtained by dilution with the same SDS solution. Solid spheres (polystyrene or poly(methyl methacrylate), (Microbeads) of different sizes (from 6 to 250 $\mu$m) are dispersed by carefully mixing them into the diluted emulsions using a spatula.
The elastic moduli of the composite materials are measured with a MCR302 rheometer (Anton Paar) by performing oscillatory strain sweeps (frequency 1 Hz, amplitude from 10$^{-5}$ to 1) in a Couette geometry with sandblasted surface. All experiments are performed at (20.0 $\pm$ 0.5) $^{\circ}$C.
 
The high concentration of surfactant used, ensures the presence of micelles and screening of electrostatic repulsion. Although the exact form of the attraction is not clear, the emulsions exhibit an elastic modulus at packing fractions below random close packing indicating the presence of an attractive interaction \cite{Datta2011}.

For the experiments with the confocal microscope the  oil was dyed with pyrromethene. The inclusion particles were not dyed so appear as black spots. The concentration of emulsion in the confocal microscopy experiment was  $\phi_{g}^* = 0.5$.\\

\section{Results}
\label{sec3}

\subsection{Numerical results}

\begin{figure}%[]
\centering
\includegraphics[width=0.45\textwidth]{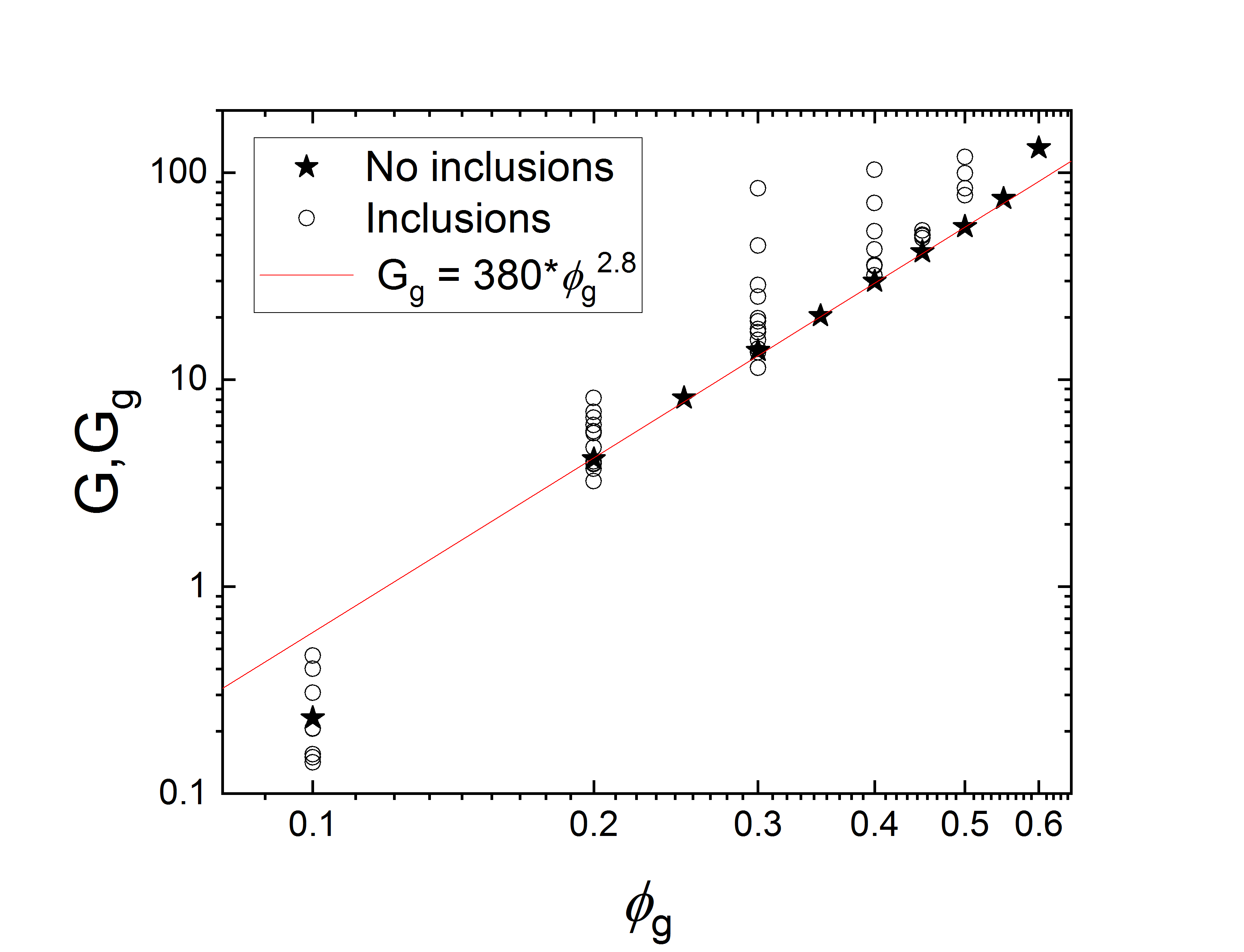}
 \caption{Simulation data for the shear elastic modulus as a function of volume fraction $\phi_{g}$ of the gel particles. Filled symbols correspond to pure gels ($G_g$) and empty symbols correspond to gels with added inclusions ($G$). 
 }
\label{fig:G0_simulations}
\end{figure}

We have measured the elastic modulus $G$ at varying gel volume fractions $\phi_{g}$ as shown in Figure \ref{fig:G0_simulations}. Our analysis focuses on the limit of small frequency, and all our $G$ values are obtained using an equivalent shear rate $\omega \tau_0=~\!1\times 10^{-3} $ and $\gamma_0=~\!1\times 10^{-4}$. The filled black data points in Figure \ref{fig:G0_simulations} correspond to samples without inclusions, pure gels, and as expected the elastic modulus of the gel increases with $\phi_{g}$. The concentrations studied in this work are high enough so that the percolating structures are not particle fractals, and models based on this assumption are indeed not suitable here. In this volume fraction range, the elasticity of the gel is no longer controlled by the individual particles but by the properties of the clusters \cite{Coussot_SoftMatter2009,Zaccone_PRL2009,RomerEPL2014}. In the range $\phi_{g} = 0.2 - 0.5$, the data are well described using a $2.8$ power law. This  power law  behaviour is close to that found in experiments by Romer et al \cite{RomerEPL2014}, where they measured 3.1 at intermediate volume fractions up to around $\phi_g$ = 0.25.

Above $\phi_g$ = 0.5, the elastic modulus is higher than predicted by the power-law, probably because we are getting deep into the glassy regime. On the other side, we observe that for our lowest density, i.e. $\phi_g =0.1$, $G_g$ is smaller than expected. This could be an indication that, for this specific concentration the fractal nature of the aggregates plays a significant role.

Next, we focus our attention on the mechanical properties of the composite gels that, as discussed before, are made by adding inclusions to the gel (different inclusion sizes $\sigma$ and volume fractions $\phi_{inc}$). The corresponding elastic modulus $G$ for a given $\phi_g$, as obtained by the computer simulations, is presented as empty symbols in Figure \ref{fig:G0_simulations}. These data points correspond to samples with size ratio values $1.0 \leq \sigma \leq 10.0$ and total packing fractions of $0.30 \leq \phi_T\leq 0.6$. We first present the general trends observed, later in the paper we will analyse those trends in more detail. In most cases the effect of the inclusions is to increase $G$. However at low gel volume fractions $G$ is shown to decrease.
 
\begin{figure}%[]
\centering
\includegraphics[width=0.45\textwidth]{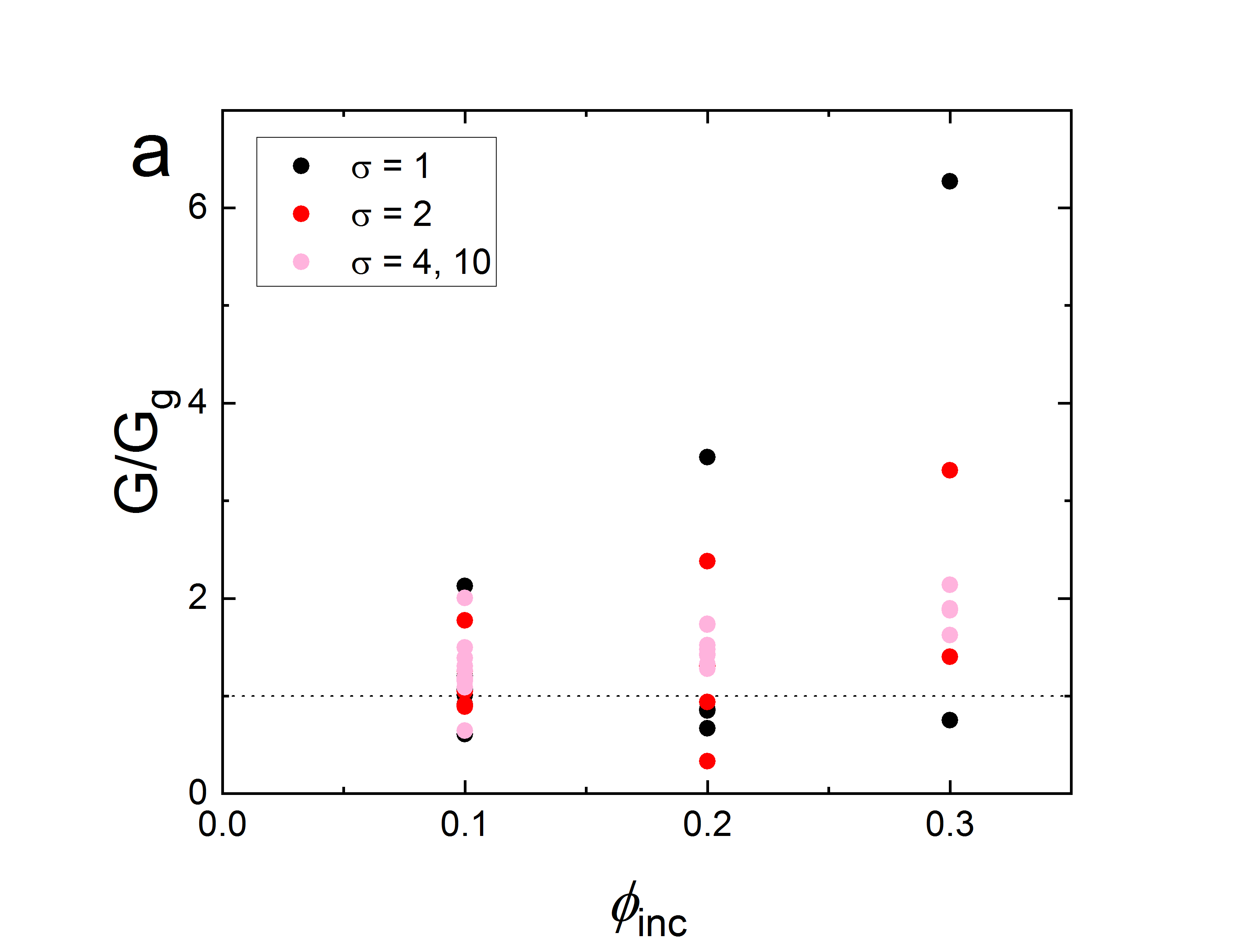}
\includegraphics[width=0.45\textwidth]{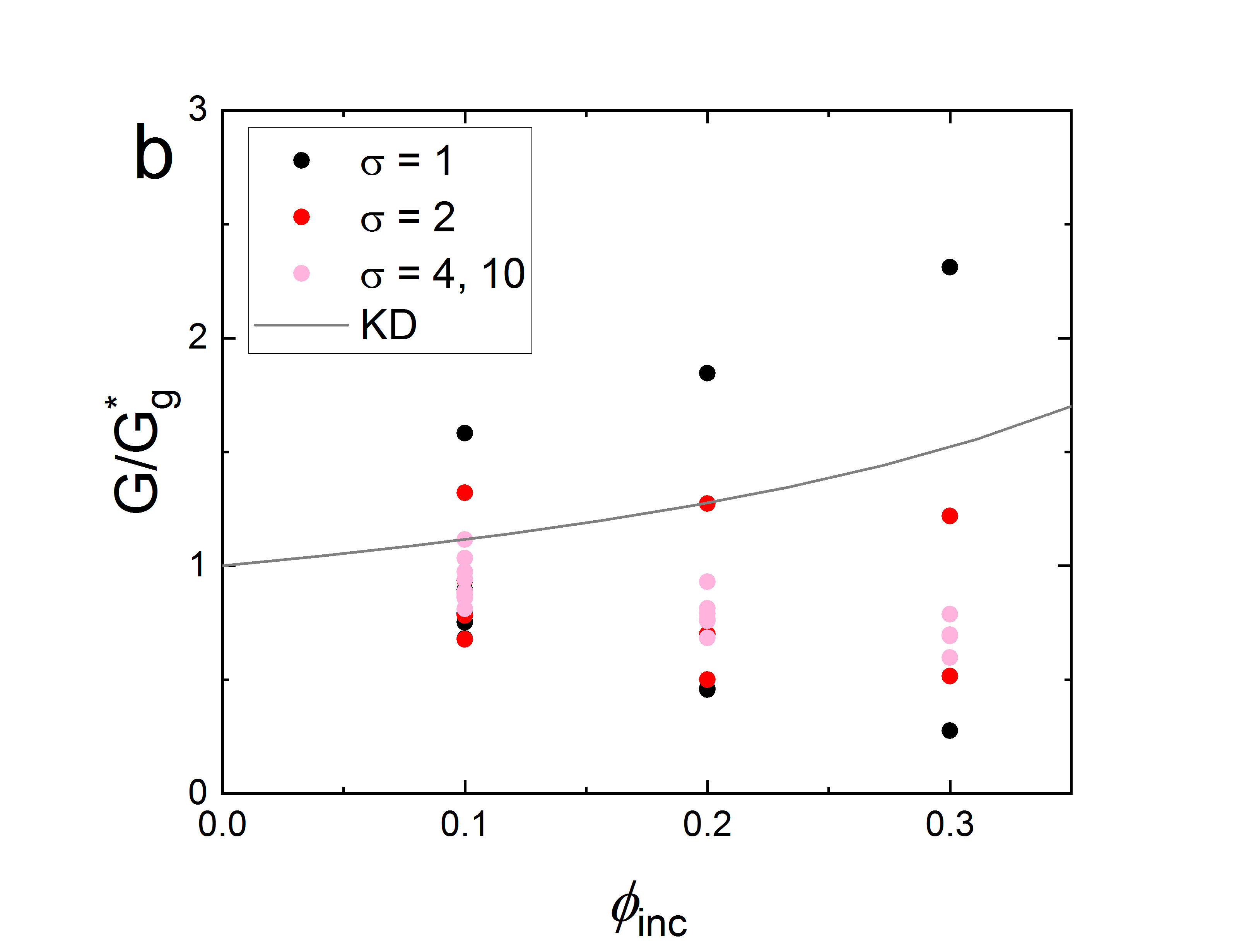}
\includegraphics[width=0.45\textwidth]{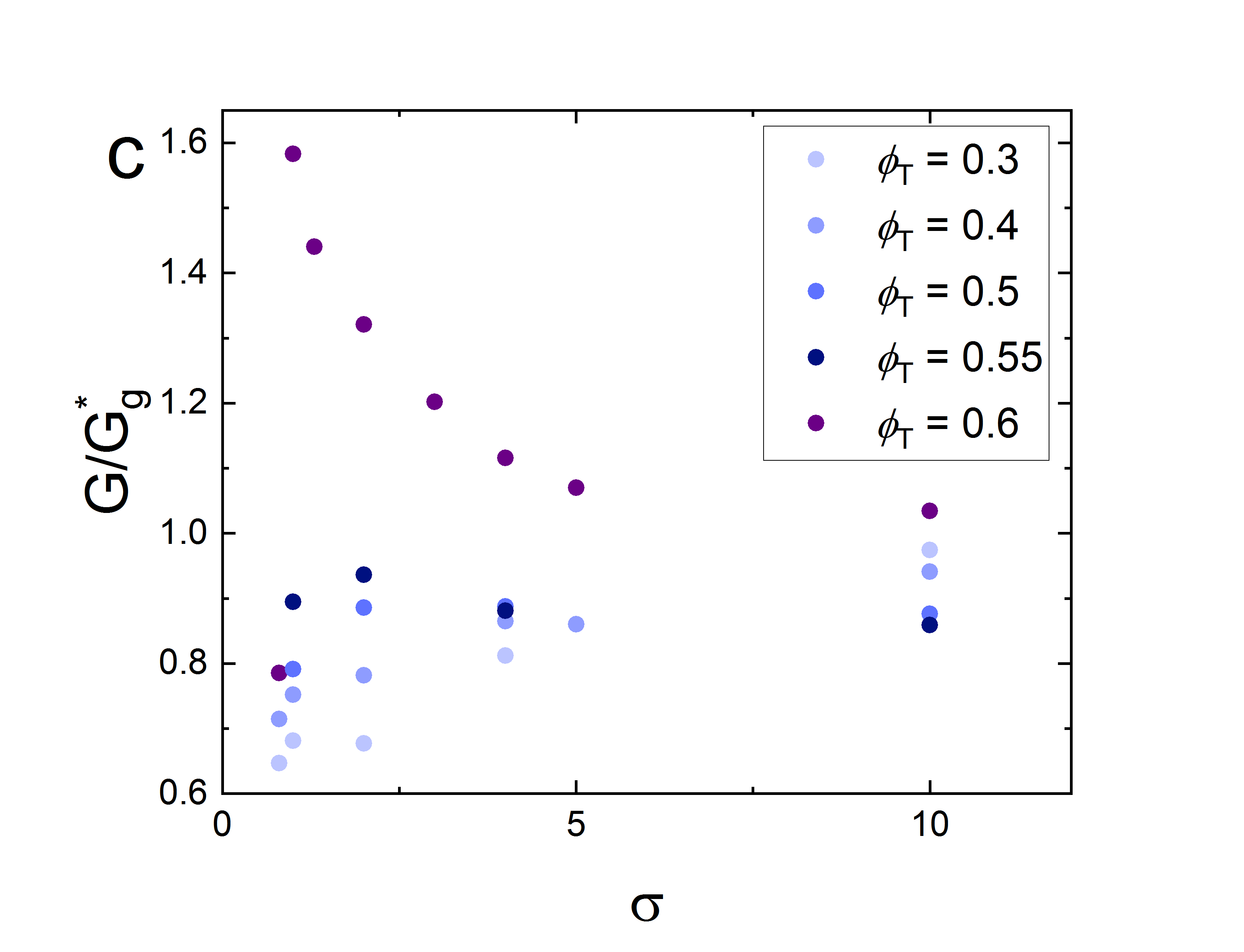}

 \caption{a) Simulation data of the elastic modulus of the composite material normalised by the elastic modulus of the bare gel shown as a function of the inclusion volume fraction. b) Same data as in a, now normalised using the gel volume fraction in the continuous phase $\phi_g^*$. The line is the KD equation with full slip boundary condition. The different inclusion sizes are shown in the legend. 
 c) The $G/G_g^*$ as a function of $\sigma$ for samples with $\phi_{inc} = 0.1$. The different $\phi_{\text T}$ are shown in the legend. }
\label{fig:G0_simulations_inclusions}
\end{figure}

To highlight the mechanical effect of the inclusions, we now plot $G/G_g(\phi_g)$ in Fig \ref{fig:G0_simulations_inclusions}a as a function of $\phi_{inc}$, the volume fraction of inclusions. The presence of the inclusions leads to a slight stiffening of the material in most cases, but two data series show a stronger increase. These are both at total volume fraction, i.e. gel particles + inclusions, $\phi_T$ = 0.6 and with the small inclusions: $\sigma$ = 1 or 2. In such cases the elastic modulus can increase at most by a factor 6.

Comparison with theory leads us to calculate also the ratio $G/G_g^*$, where the elastic modulus of the matrix $G_g^*$ is described by $G_g^* = G_g(\phi_g^*)$ using $\phi_g^* = \phi_{g}/(1-\phi_{inc})$, i.e. the volume fraction of the gel embedding the inclusions. This normalisation is shown in Fig \ref{fig:G0_simulations_inclusions}b. Again, we observe two different behaviours depending on the total volume fraction $\phi_{\text T}$. For $\phi_{\text T} = 0.6$, the effect of the inclusions is to stiffen the composite gel with a magnitude that depends strongly on the size ratio $\sigma$: the smaller is $\sigma$ the more pronounced is the variation, reaching a maximum of roughly $2$ times the elastic modulus of the pure gel.

As a theoretical reference we will consider the Krieger and Dougherty (KD) equation \cite{KriegerDougherty_1959}, which was originally proposed to describe the shear viscosity of non-Brownian solid particles in Newtonian fluids \cite{Larson_1999}. Note that the problem of elasticity of rigid particles embedded with a linear elastic material is formally similar to the problem of the viscosity of rigid particles suspended in a Newtonian liquid. In the KD equation the elastic modulus increases with the volume fraction of solid inclusions $\phi_{inc}$, as:
\begin{equation}
    \frac{G}{G_g^*} = \left( 1-\frac{\phi_{inc}}{\phi_m} \right) ^{-x \phi_m}
\label{KD_equation}
\end{equation}
with $\phi_m$ the close packing volume fraction at which the elastic modulus diverges and where the coefficient $x = 2.5$ has been conveniently introduced to recover the Einstein equation \cite{Einstein_1906} at low concentration (first order in $\phi_{inc}$), i.e. $G/G_g^* \simeq 1+\frac{5}{2} \phi_{inc}$. For spherical inclusions characterized by slip condition at their surface, just like the inclusions we have simulated, the equation derived by Taylor \cite{Taylor_1932,F&A_1970}, i.e. $G/G_g^* \simeq 1+ \phi_{inc}$, is more appropriate at low volume fraction. For higher volume fractions, empirical equations were proposed \cite{Pal_1992} but, for the sake of simplicity, it is sufficient to use Eq. \ref{KD_equation} with $x = 1$ for comparison with our simulations. Such a comparison is presented in Figure \ref{fig:G0_simulations_inclusions}b, using $\phi_{\text m} = 0.59$. The elastic modulus is below the KD prediction as a function of $\phi_{inc}$ for all the samples except for the $\sigma = 1$ at $\phi_{\text T} = 0.6$.

To discuss further the results, we present $G/G_g^*$ in Fig.~\ref{fig:G0_simulations_inclusions}c as a function of $\sigma$ for a fixed $\phi_{inc} = 0.1$ but varying the gel concentration $\phi_g$. When the total volume fraction is below $0.6$, the elastic modulus displays a weak or moderate dependence on $\sigma$, essentially a decrease. On the other side, when the total volume fraction reaches $0.6$, the system does display a remarkable sensitivity to the size of the inclusions, with an up to one and a half times larger value of the elastic modulus with respect to the pure gel. We can conclude that the smaller are the inclusions the stronger is the effect on the mechanical properties, a behaviour whose origin can be understood by looking at the microscopic interplay between the gel and the inclusions.

\begin{figure}%[]
\centering

\includegraphics[width=0.55\textwidth]{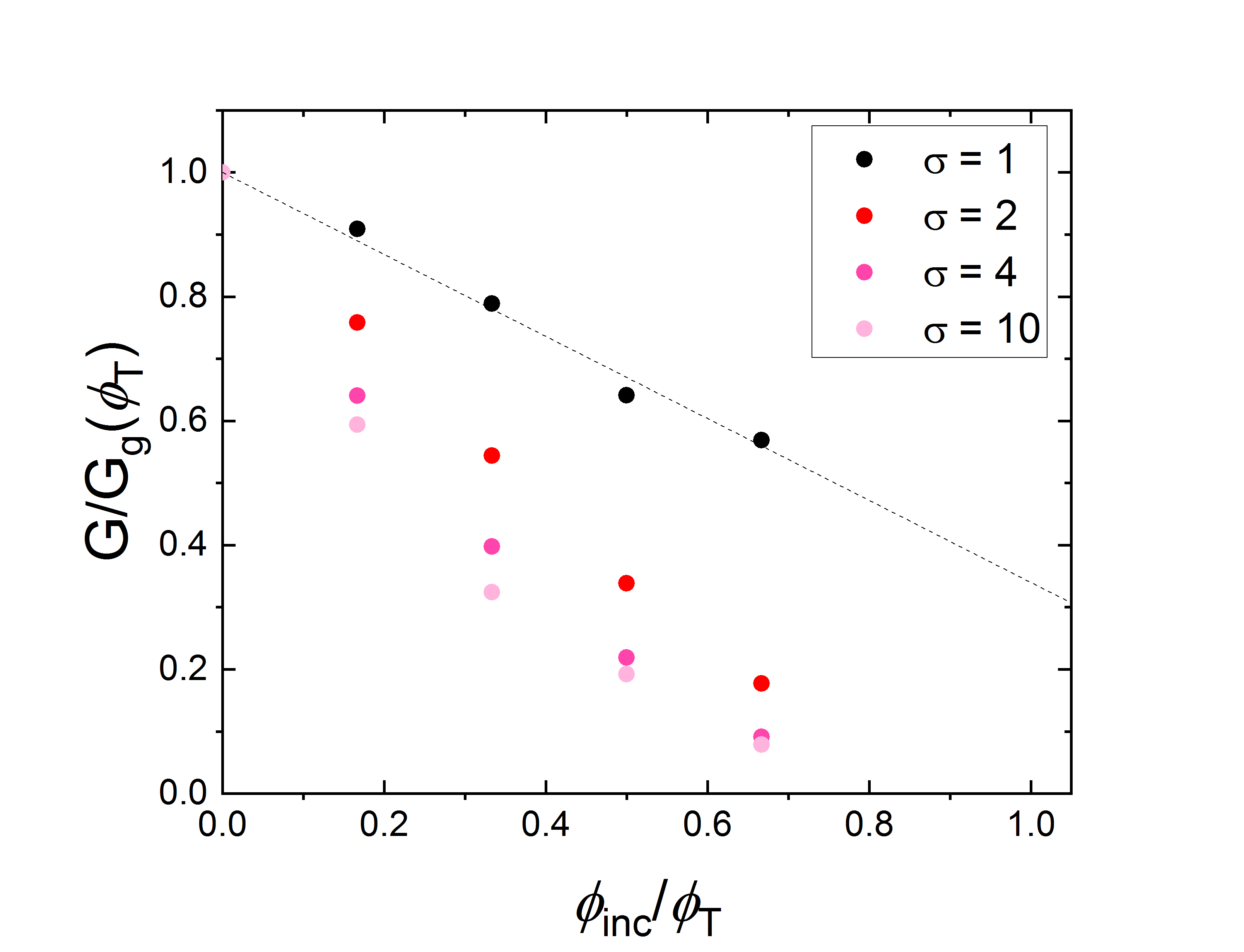}
 \caption{ Elastic modulus of composite gel as a function of the fraction of particles which are inclusion particles instead of gel, i.e. $\phi_{inc}/ \phi_{T} = 0$ refers to a sample with no inclusions and $\phi_g = 0.6$.  In all the samples shown $\phi_{\text T}$ = 0.6. }
\label{fig:Interaction_test}
\end{figure}

In such a granular (colloidal) system, the elastic modulus depends on the number density of contacts, the stiffness of the contact and the gel structure (texture). By considering a dense homogeneous gel at total volume fraction $\phi_{\text T} = 0.6$, we can assume that the structure is not significantly modified if gel particles are replaced by inclusions of the same size. In this case, we could write $G_g(\phi_{\text T}) \propto z\phi_T \chi_{g-g} / \langle R_g \rangle $, where $z$ is the coordination number and $\chi_{g-g}$ is the contact stiffness between two gel particles\cite{Zaccone_PRL2009}. In the composite gel at $\phi_{\text T}$, volume fraction $\phi_{inc}$ of gel particles have been replaced by non-attractive inclusions, and this leads to a decrease of the modulus. With those assumptions, the corresponding modulus writes $G \propto \frac{z\phi_{\text T}}{\langle R_g \rangle} \left (\frac{\phi_g}{\phi_{\text T}}\chi_{g-g} + \frac{\phi_{inc}}{\phi_{\text T}}\chi_{g-i} \right)$, where $\chi_{g-i}$ represents the effective contact stiffness between gel particles and inclusion particles. Therefore, the reduced elastic modulus can be expressed as:
\begin{equation}
    \frac{G}{G_g(\phi_{\text T})}  = 1 - \frac{\phi_{inc}}{\phi_{\text T}} \left (1- \frac{\chi_{g-i}}{\chi_{g-g}} \right).
\label{microequation}
\end{equation}
A single parameter fit to the data, shown by the dashed line in Fig.~\ref{fig:Interaction_test}, yields $\frac{\chi_{g-i}}{\chi_{g-g}}$ = 0.34. This shows how the replacement of $g-g$ interaction by $g-i$ interaction contributes to decrease the shear elastic modulus of those loaded particle gels. As can be seen in Fig.~\ref{fig:Interaction_test}, the effect of size ratio $\sigma$ is to decrease the elastic modulus with respect to Eq. \ref{microequation}. This behavior is expected to be partly related to the well-known increase of the maximum volume fraction of bimodal granular assemblies \cite{Farr_bidisperse}. Note that here we consider that $\phi_{\text T}$ is constant, so the volume fraction of gel particles between the inclusions $\phi_g^* = (\phi_{\text T} - \phi_{inc}) / (1 - \phi_{inc})$ decreases significantly as a function of $\phi_{inc}$.

\begin{figure}%[]
\centering
\includegraphics[width=0.45\textwidth]{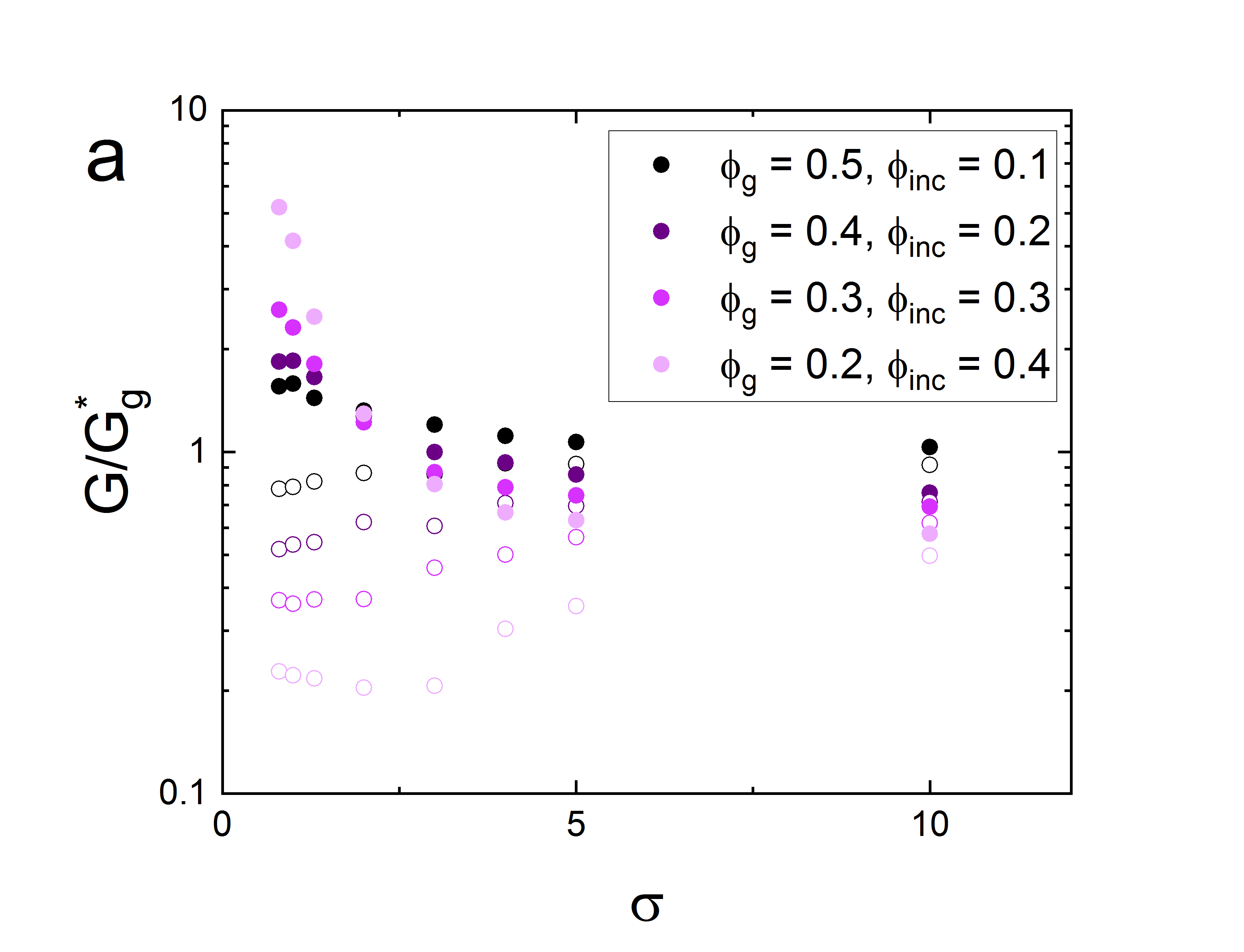}
\includegraphics[width=0.45\textwidth]{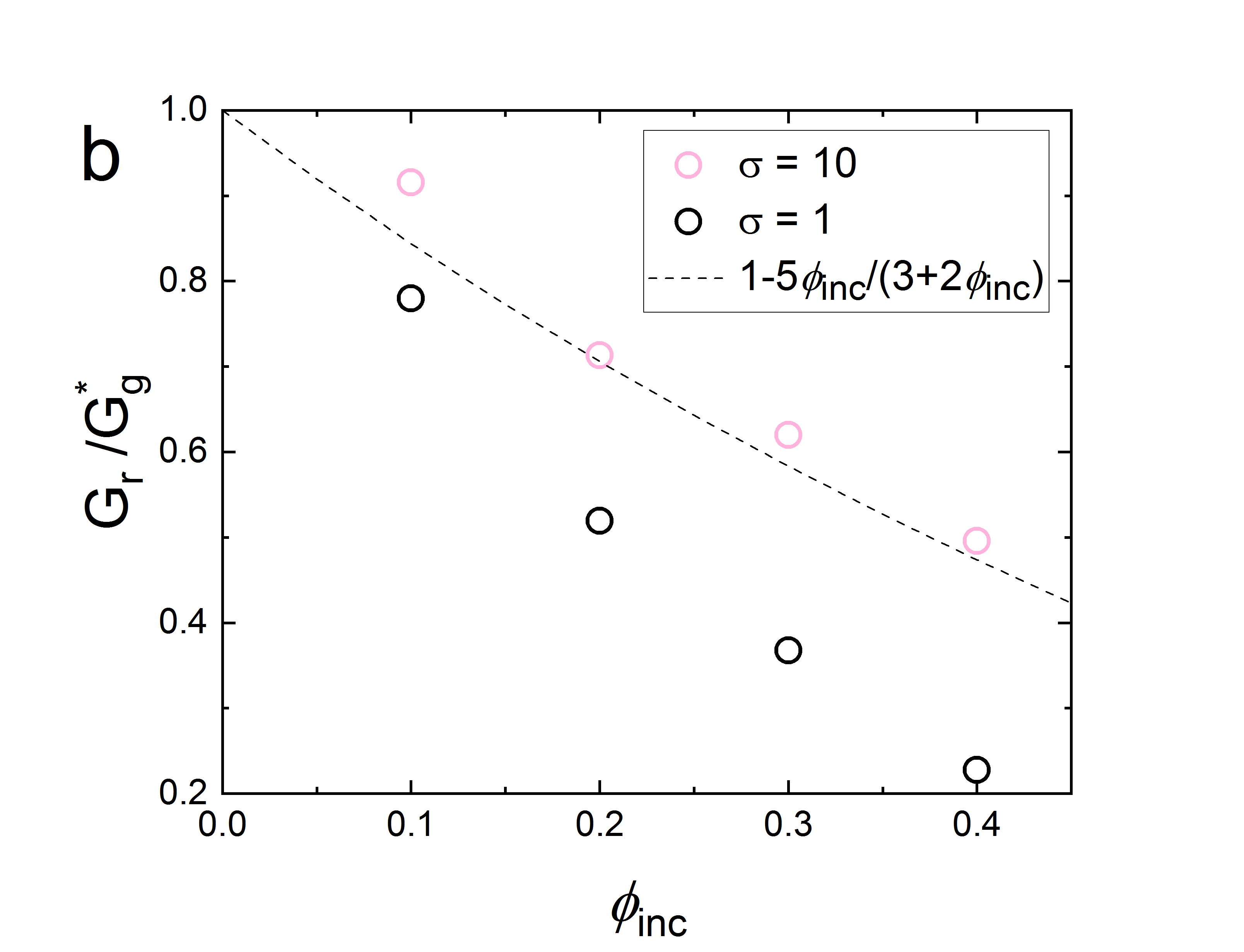}
\includegraphics[width=0.45\textwidth]{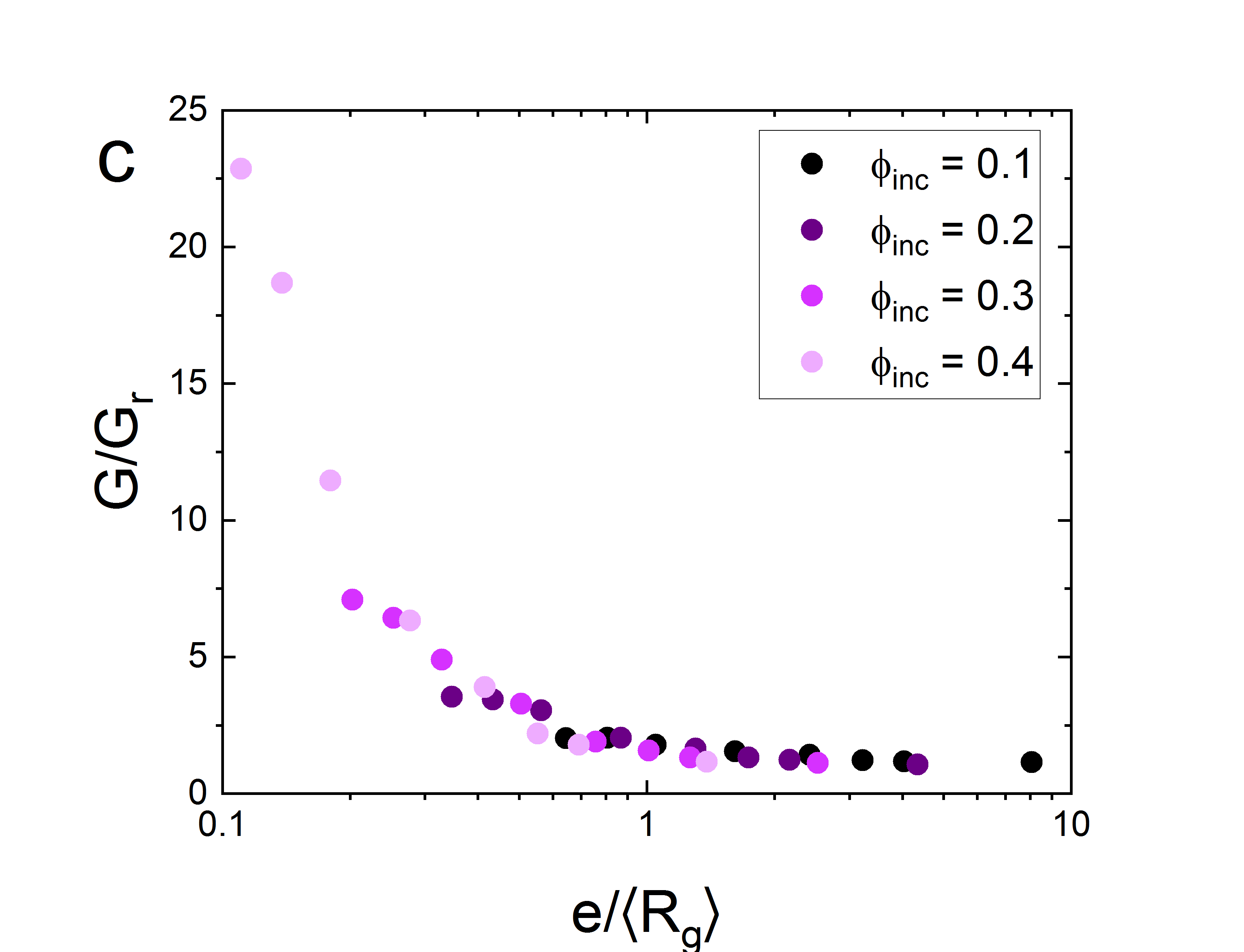}
\caption{a) Elastic modulus of the composite material normalised by the elastic modulus of the continuous phase as a function of the size ratio of the inclusions. The filled circles correspond to systems with the solid inclusions present, the empty circles are the same systems, but once the solid inclusions are removed. 
 b) Normalised elastic modulus of the samples with inclusions removed and the prediction of the effect of added holes.  
 c) Elastic modulus of the composite normalised with the elastic modulus once the inclusions are removed as a function of the distance between inclusion particles. }
\label{fig:takingoutinclusions}
\end{figure}

\begin{figure}%[]
\centering
\includegraphics[width=0.45\textwidth]{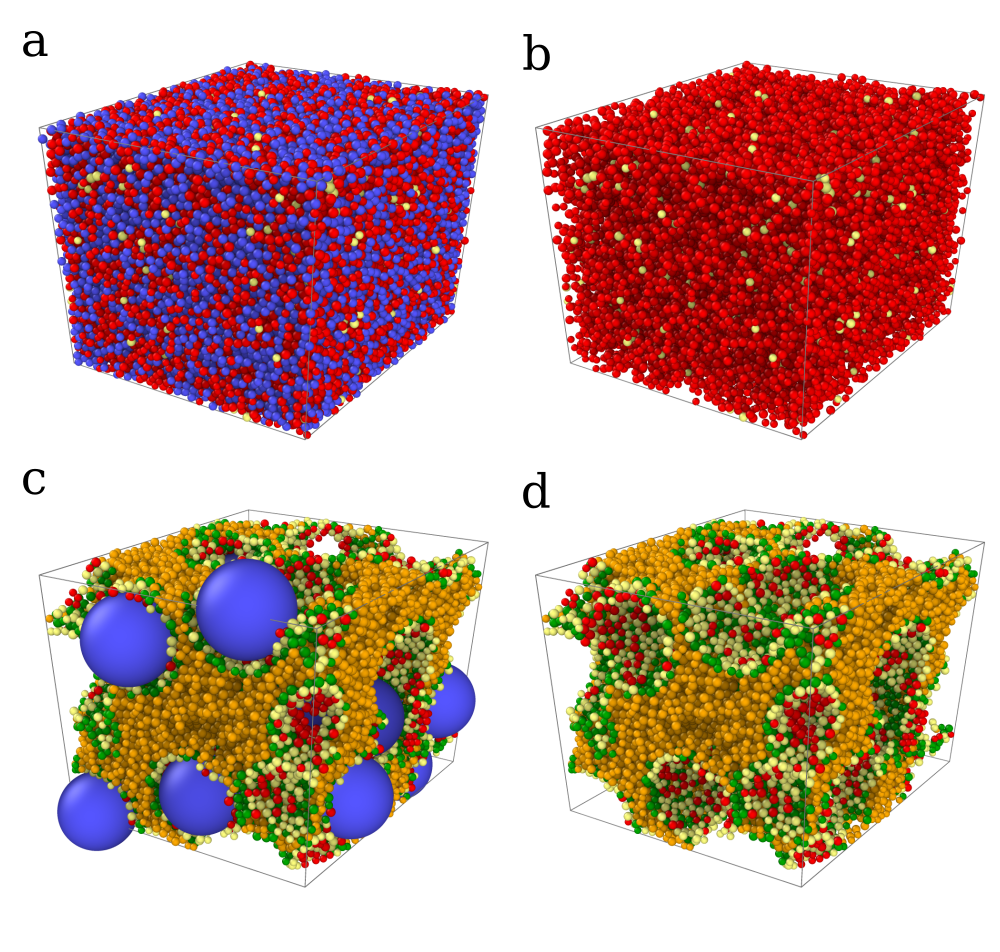}

\caption{Snapshots of the simulations a) and b) are with $\sigma$ = 1 and c) and d) with $\sigma$ = 10. In a and c inclusion particles are coloured in blue while the gel particles are coloured  according to their proximity to an inclusion. If their distance from an inclusion is less than $D_{min}$ (the cut-off of the potential) the particles are coloured in red; all other colours correspond to bigger separation distances. In b and d the inclusions have been removed.  }
\label{fig:snap_his}
\end{figure}

More drastic changes can be anticipated in the gel structure due to the introduction of inclusions. As an example, and with contrast to continuous matrices, gel particles can be excluded from the gaps between inclusions. To highlight such structure effect, we performed mechanical tests on samples where we removed the inclusions. In this protocol, the gel assembles in the presence of inclusions, but the latter are then removed before proceeding to the mechanical test. In Fig.~\ref{fig:takingoutinclusions}a, we compare the system with and without the inclusions in a similar fashion to what was done in Fig.~\ref{fig:G0_simulations_inclusions}, but here only for $\phi_{\text T} =0.6$ where the size effects appeared to be most prominent. It is shown that the dependence on $\sigma$ is much weaker than the one observed for systems including the inclusions. At this stage it is interesting to compare these results with the theory for continuous matrix with holes, as given by micromechanics calculation \cite{F&A_1970,Dormieux_2006}: $G_r/G_g^* = 1-5\phi/(3+2\phi)$, where $\phi$ is the volume fraction of holes (here we have $\phi = \phi_{inc}$). Comparison is shown in Fig~\ref{fig:takingoutinclusions}b for two size ratios ($\sigma = 10$ and $\sigma = 1$). It appears that results for $\sigma = 10$ are compatible with the theory, which can be understood by the fact that $\phi_g^*$ is high enough for the gel particle assembly to be shaped by the precursor inclusions as a thick skeleton (see Fig.~\ref{fig:snap_his}). In contrast, for $\sigma = 1$ such a skeleton does not exist because the precursor inclusions have altered the gel particle network (see Fig.~\ref{fig:snap_his}) as numerous weakness points. As a result, the second configuration is weaker than the first one although gel volume fractions are the same.

$G_r$ appears to be the relevant elastic modulus to be considered when quantifying the effects of the inclusions, so we divided the composite gel modulus $G$ by its modulus with removed inclusions, $G_r$. Moreover, one expects finite size of the gel particles to come into play when the mean gap size $e$ between inclusions is close or smaller than the gel particle size, i.e. $2\langle R_g \rangle$. According to the parameters defined above, the ratio $e/2\langle R_g \rangle$ writes: $\sigma((\frac{\phi_m}{\phi_{inc}})^{\frac{1}{3}}-1)$. The resulting values are used to plot $G/G_r$ against $e/2\langle R_g \rangle$ in Fig.~\ref{fig:takingoutinclusions}c, where a fair collapse is obtained for all the different cases. This indicates that the two contributions, i.e. interaction and structure effects, can be decoupled for this total volume fraction and, once the change in the backbone structure has been taken into account, effects of interactions of the inclusions with the gel particles can be described by the ratio $e/2\langle R_g \rangle$. It appears that although the effective stiffness is weaker for interactions involving inclusions, small inclusions (i.e. $e/2 \langle R_g \rangle < 1$) are embedded within the gel particle backbone and provide strengthening by supporting the backbone. On the other hand, larger inclusions (i.e. $e/2\langle R_g \rangle \gg 1$) are not included into the backbone but rather shape it around them, providing strengthening comparable to prediction of the classical theory, i.e. KD. 

\begin{figure}%[]
\centering

\includegraphics[width=0.45\textwidth]{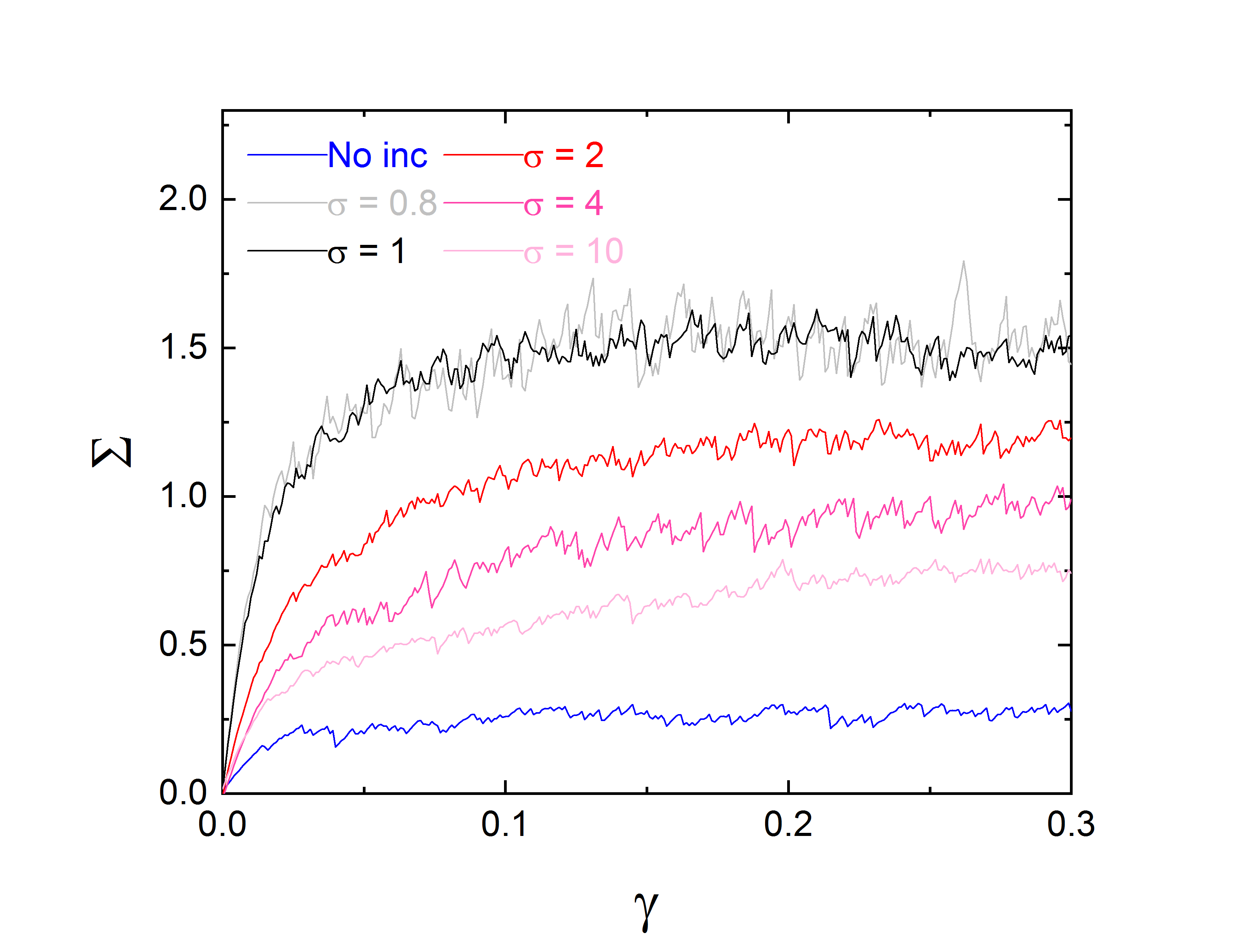}
 \caption{Stress-strain curves for samples with $\phi_g$ = 0.3 without inclusions (blue curve) and with $\phi_{inc}$ = 0.3 for different $\sigma$. Shear rate $\dot{\gamma}\!~=~\!10^{-5}\tau_0^{-1}$ and strain step $\Delta \gamma\!~=~\!0.001$. }
\label{fig:Continuos_strain}
\end{figure}

To pin down the microscopic origin of the mechanical response discussed above, we performed continuous strain deformation simulations to unveil the microscopic stress experienced by each particle in our composite gels.
We focus on the case of $\phi_{inc} = 0.30$ and $\phi_{g} = 0.30$ and we simulated, using step-strain deformation, a continuous deformation in the $xy$ plane. The resulting stress-strain curves for different inclusion sizes as well as for the pure gel case are displayed in Fig.~\ref{fig:Continuos_strain}. We immediately notice a strong difference between the response of the pure gel and the composite gel, however this does not come as a surprise given the fact that for the pure gel there are no inclusions that reinforce the overall structure. A more relevant comparison is for systems with different size of inclusions and, in this case, several different trends are observed. In agreement with the measurement of the elastic modulus discussed above, for small strain the slope of the curve grows when the size of the inclusions decreases. For large inclusions, we notice that, upon straining, stress builds up continuously. For small inclusions, in particular when their size is similar to the size of the gel, the strain-stress curve presents clear signs of yielding. In this case, at $\gamma\simeq 0.10$ we notice that the curve has leveled off and reached a value that is more than three times larger than the stress observed for larger inclusion sizes, which confirms the differences that we have observed before.

\begin{figure}%[]
\centering
\includegraphics[width=0.45\textwidth]{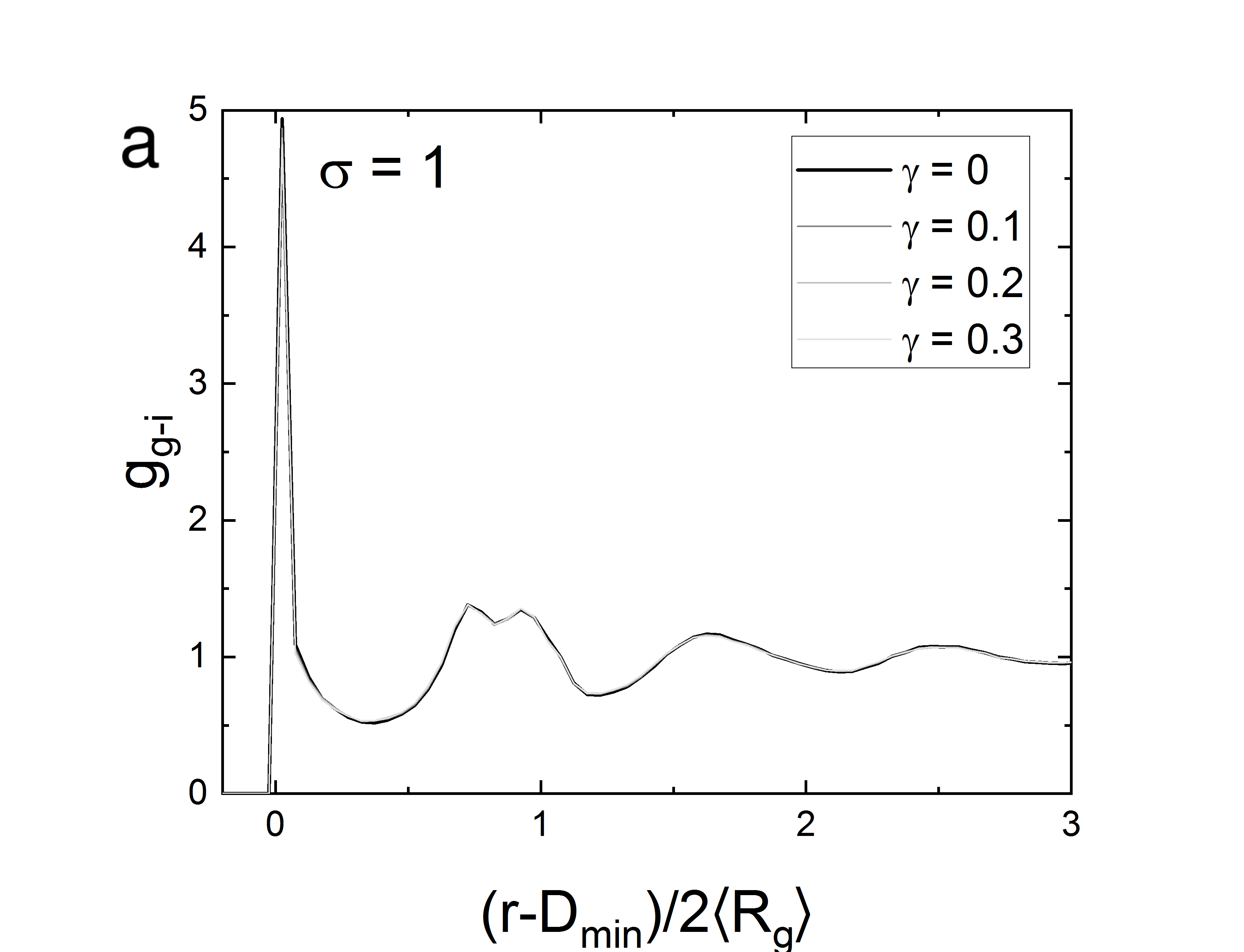}
\includegraphics[width=0.45\textwidth]{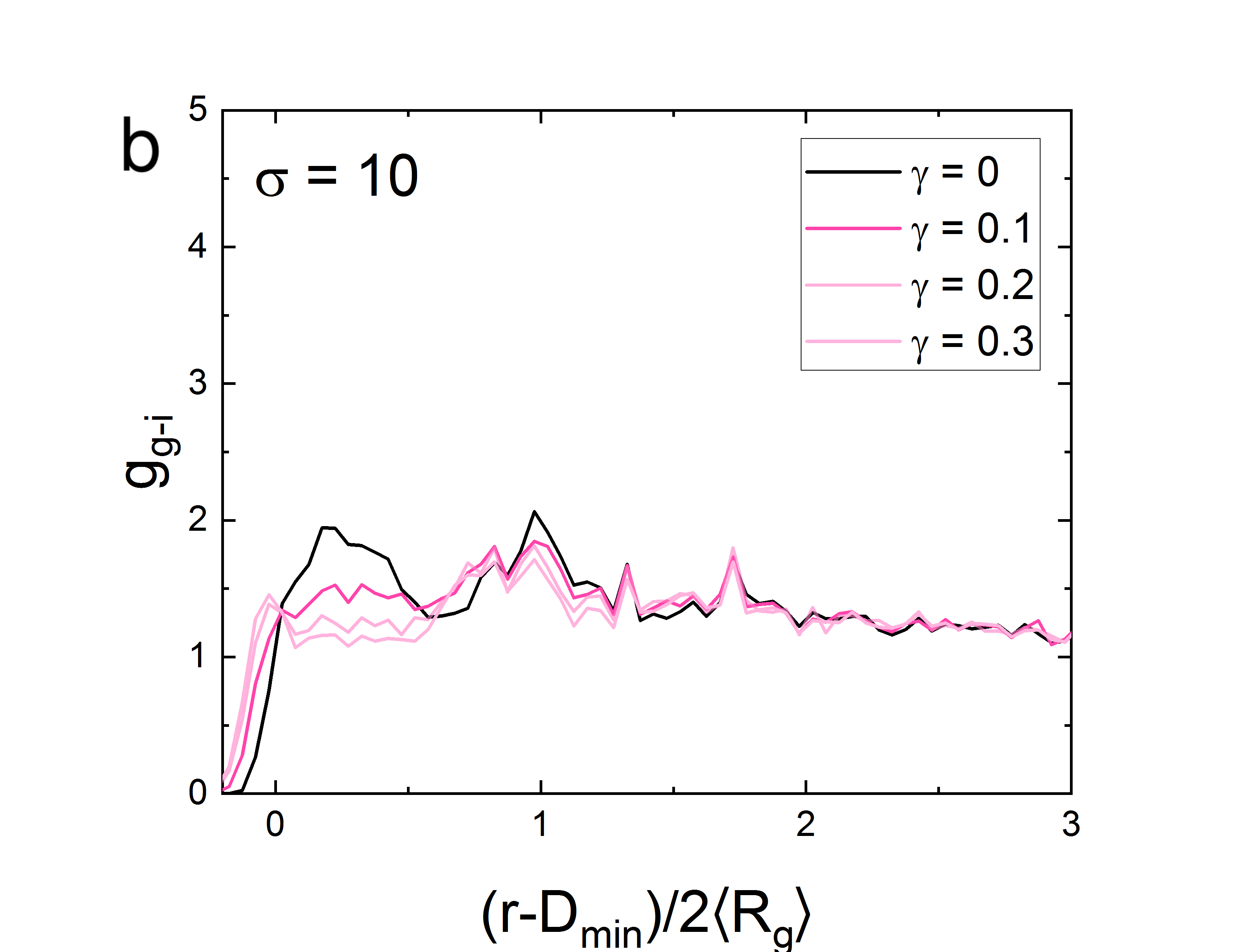}

 \caption{Gel-inclusion particle radial distribution function as a function of deformation. The results are for two representative size ratios $\sigma=1$ (a) and $\sigma=10$ (b).}
\label{fig:rdf}
\end{figure}

Next, we move to the discussion of the structural properties of the composite gel and, from now on, we will focus only on two representative cases of size ratios, namely $\sigma=1$ and $\sigma=10$.  In Fig.~\ref{fig:rdf} (a) and (b), the inclusion-gel radial distribution functions are presented as a function of the non-dimensional distance $(r-D_{min})/2\langle R_{g} \rangle$, where $D_{min}$ is the cut-off distance after which there is no net force between two particles and $2\langle R_{g} \rangle$ is  the average diameter of the gel particles. A remarkable difference appears between the two cases, despite the fact that the $\sigma=10$ curves present some noise due to finite size. For the large size ratio, the  distribution is almost flat while, for smaller size ratio, there is a pronounced peak. This is an important information because only the particles  for which $(r-D_{min})/2\langle R_{g} \rangle < 0$ will exert a net force on each other.\\ 

\begin{figure}%[]
\centering
\includegraphics[width=0.45\textwidth]{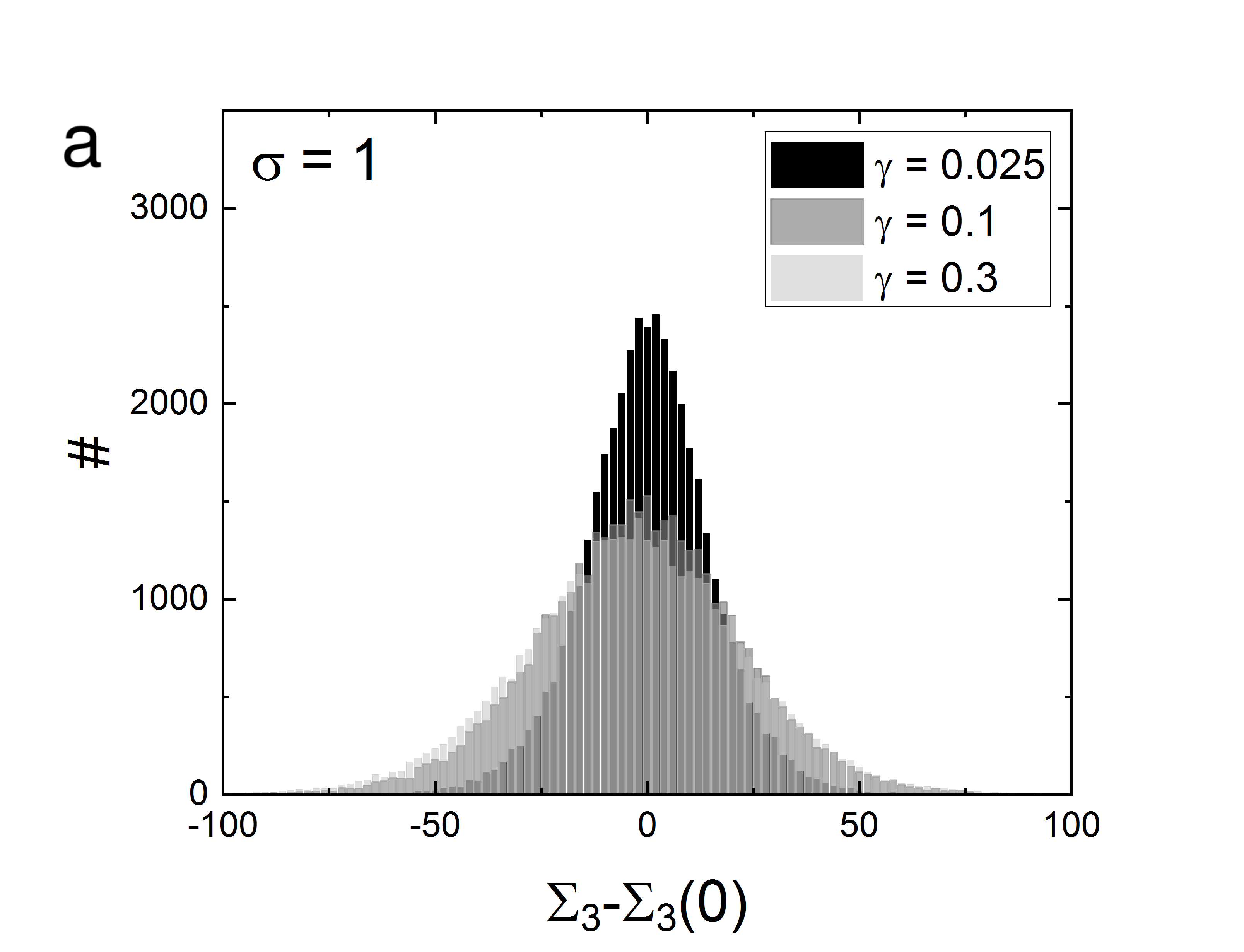}
\includegraphics[width=0.45\textwidth]{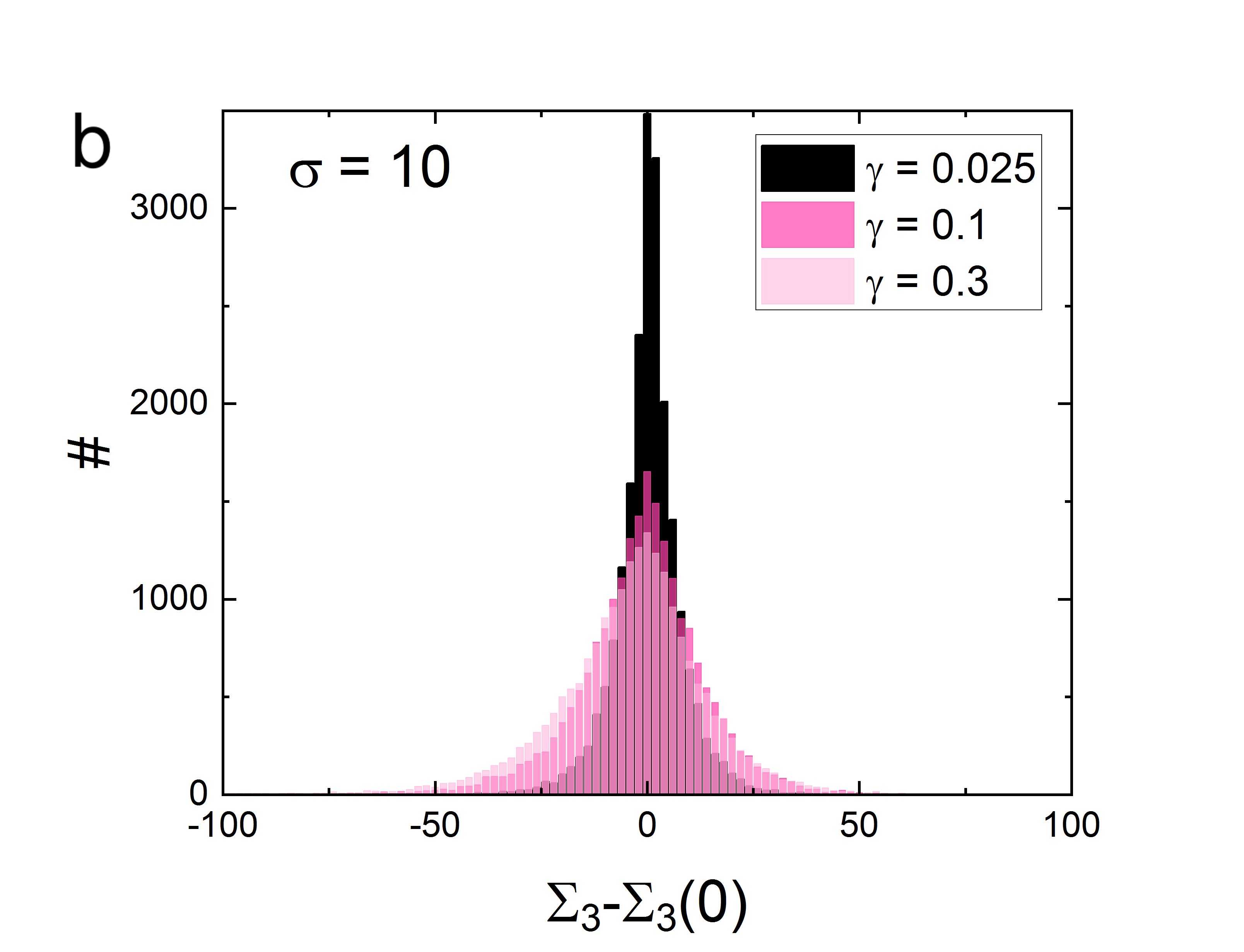}
\includegraphics[width=0.45\textwidth]{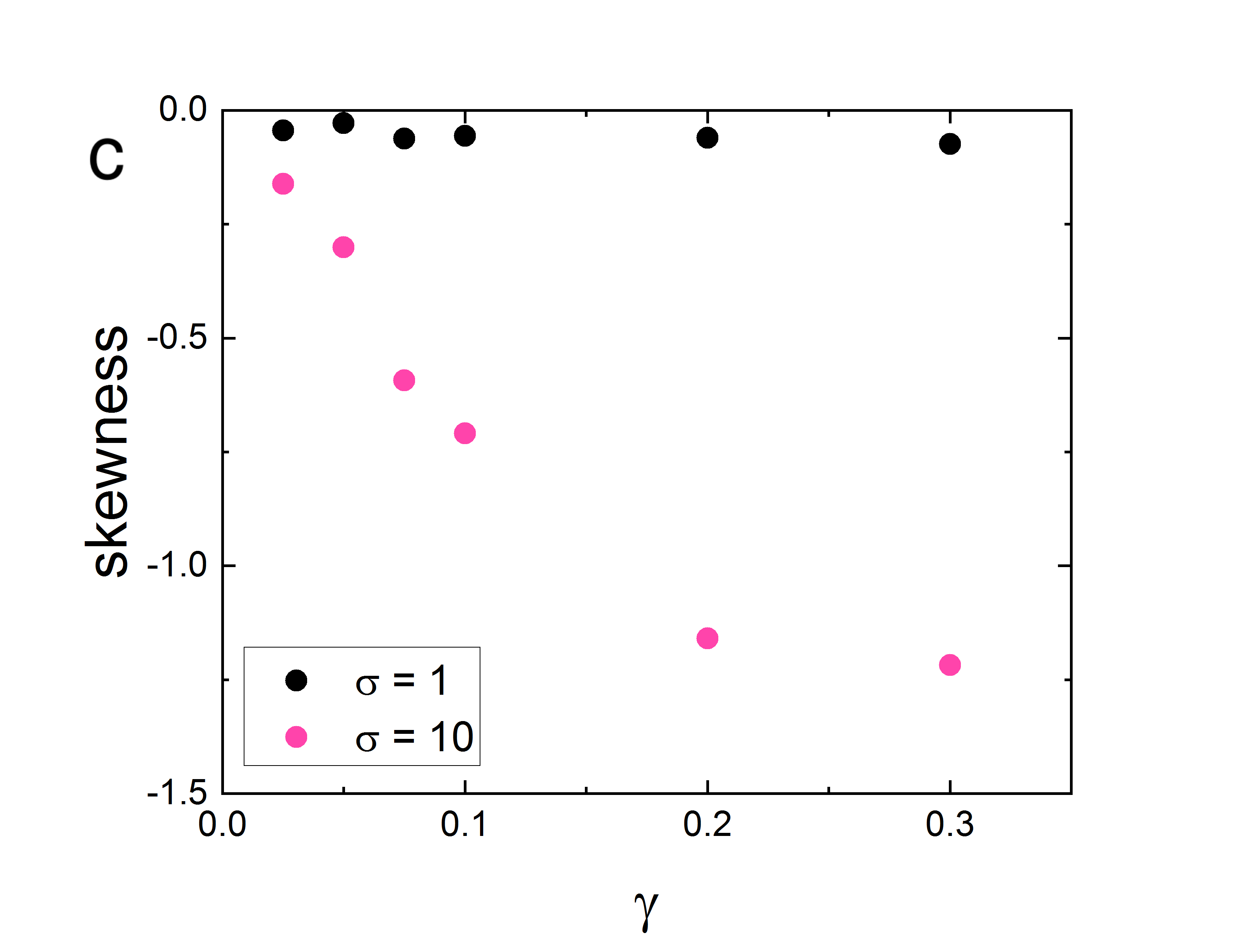}

 \caption{Local stress histogram distribution at different strain for two representative size ratios $\sigma=1$ (a) and $\sigma=10$ (b). In panel c, we present variation of the skewness of the two distributions as a function of the strain $\gamma$. 
 }
\label{fig:histo_interactions}
\end{figure}

A more direct way to test this observation is to perform a direct calculation of the local stress acting on the gel backbone. To this aim, we  calculated the local stress tensor $\Sigma^i$ for each particle $i$ and its three eigenvalues $\Sigma_1^i \geq \Sigma_2^i  \geq \Sigma_3^i$ which represent the three invariant principal stresses in the frame of reference of the stress. This stress tensor has been normalized with the volume of the corresponding particle. We focus on $\Sigma_3^i$, which indicates the largest compression on each particle, and from now on we will refer to it as $\Sigma_3$. Note that this value is not the same as the shear component $\Sigma$ used in previous sections. In the calculation of the stress tensor for each particle, all the interactions between particles are taken into account, this includes gel-gel, gel-inclusion and inclusion-inclusion interactions. For different strain values, we have tracked the variation of $\Sigma_3$ with respect to the undeformed case $\gamma=0$, i.e. $\Sigma_3 - \Sigma_3(0)$. The histograms for this quantity, for  $\sigma=1$ and $\sigma=10$, are presented in Fig.~\ref{fig:histo_interactions}a and Fig.~\ref{fig:histo_interactions}b respectively for different values of $\gamma$. In Fig.~\ref{fig:histo_interactions}c we calculate the skewness of the two distributions as a function of the strain. In the case $\sigma=1$, the distribution remains basically symmetric around the minima whereas, for $\sigma=10$, the distribution becomes remarkably skewed toward the negative values.

This trend is clearly related to the structural properties of the composite gel and, in Fig.~\ref{fig:snap_his}, we present snapshots of the same systems for $\gamma=0.3$. Inclusion particles are coloured in blue while the gel particles are coloured  according to their proximity to an inclusion. In particular, if their distance from an inclusion is less than $D_{min}$ (the cut-off of the potential) the particles are coloured in red. It is evident that, for $\sigma=10$, the number of particles in close proximity to  an inclusion is much smaller than for the case $\sigma=1$. In the second case almost all particles in the gel are in contact with at least one inclusion. The inclusions reinforce the system and help reduce compression in comparison with the pure system (or the composite mixture with $\sigma=10$).
We conclude that the unexpected trend of the mechanical response of our composite gel is the result of a subtle interplay between the structural properties and the mechanical interactions between the inclusions and the gel and for this reason it depends strongly on the size ratio between the two. 

\subsection{Experimental results}

As explained in the Sec~\ref{Sec: Meth_Exp}, we explored the influence of the matrix grain size using oil-in-water emulsions. Those emulsions are made of polydisperse oil droplets with an average size of around 5 $\mu$m. Several common points with our numerical system can be noted as follows: (1) the droplets can be well-approximated by spheres for the oil fractions we studied, and (2) the drops are stabilized using an excess of ionic surfactant, which is at the origin of the attractive interaction between the drops (depletion forces). With respect to the second point, note that electrostatics just plays a marginal role in the interaction since it is screened due to the high concentration of surfactant counter-ions.

On the other hand, our experimental system could appear to be very different from the ideal numerical system due to the intrinsic softness of the oil droplets. Actually, we think this difference is an interesting point to investigate. Indeed, the addition of solid spheres (inclusions) in oil-in-water emulsions has been shown to increase the elastic modulus according to the KD model \cite{Mahaut2008, Vu_2010}. But this was done only in the limit where the size of the inclusions is much bigger than the characteristic size in the matrix ($\sigma \gg 1$). Therefore this system is of particular interest to seek for the finite size effects highlighted in the previous section. Besides, note that our results have been obtained out of the size ratio range corresponding to $\sigma \ll 1$, for which the inclusion particles are allowed to organize between the particles of the matrix in the shape of a skeleton-like packing, providing strongly amplified strengthening \cite{Gorlier_PRE_2017,Gorlier_SoftMatter_2017,Addad_PRL_2007}.

%- snapshots

\begin{figure}
\centering
\includegraphics[width=0.45\textwidth]{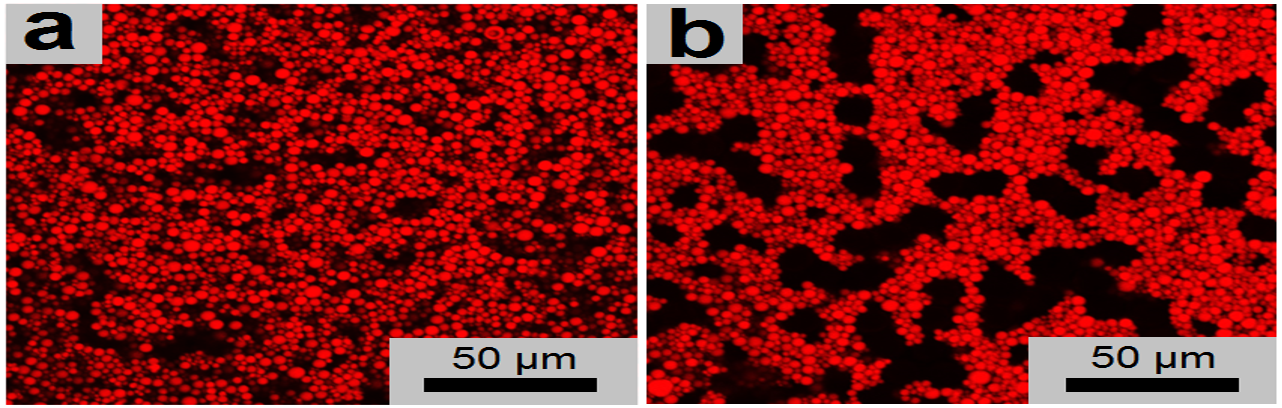}
 \caption{a) Image taken using a confocal microscope of an emulsion at $ \phi_{g}^* = 0.5$ without inclusions and  b) The same emulsions with added inclusions with $\sigma = 2.5$, resulting in  $\phi_g^* = 0.5$ and $\phi_{inc} = 0.1$.}
\label{fig:Confocal_experiments}
\end{figure}

A microscope image of an emulsion without inclusions with $\phi_g = 0.5$ is shown in Figure \ref{fig:Confocal_experiments}a. The emulsion drops have been dyed using pyrromethene, which fluoresces in red. In this attractive system, as we are below close packing of the emulsions, the system will present a gel-like structure with its typical heterogenous structure. The size of the pores will depend on the attraction as well as on the oil volume fraction. Once added to the system, the inclusions will sit in these cavities and, as a consequence, will profoundly affect the structure of the gel matrix. An example of the resulting composite material  is shown in Figure \ref{fig:Confocal_experiments}b with $\phi_{inc} = 0.1$ and with $\sigma = 2.5$. 

%- rheology results description

\begin{figure}
\centering
\includegraphics[width=0.45\textwidth]{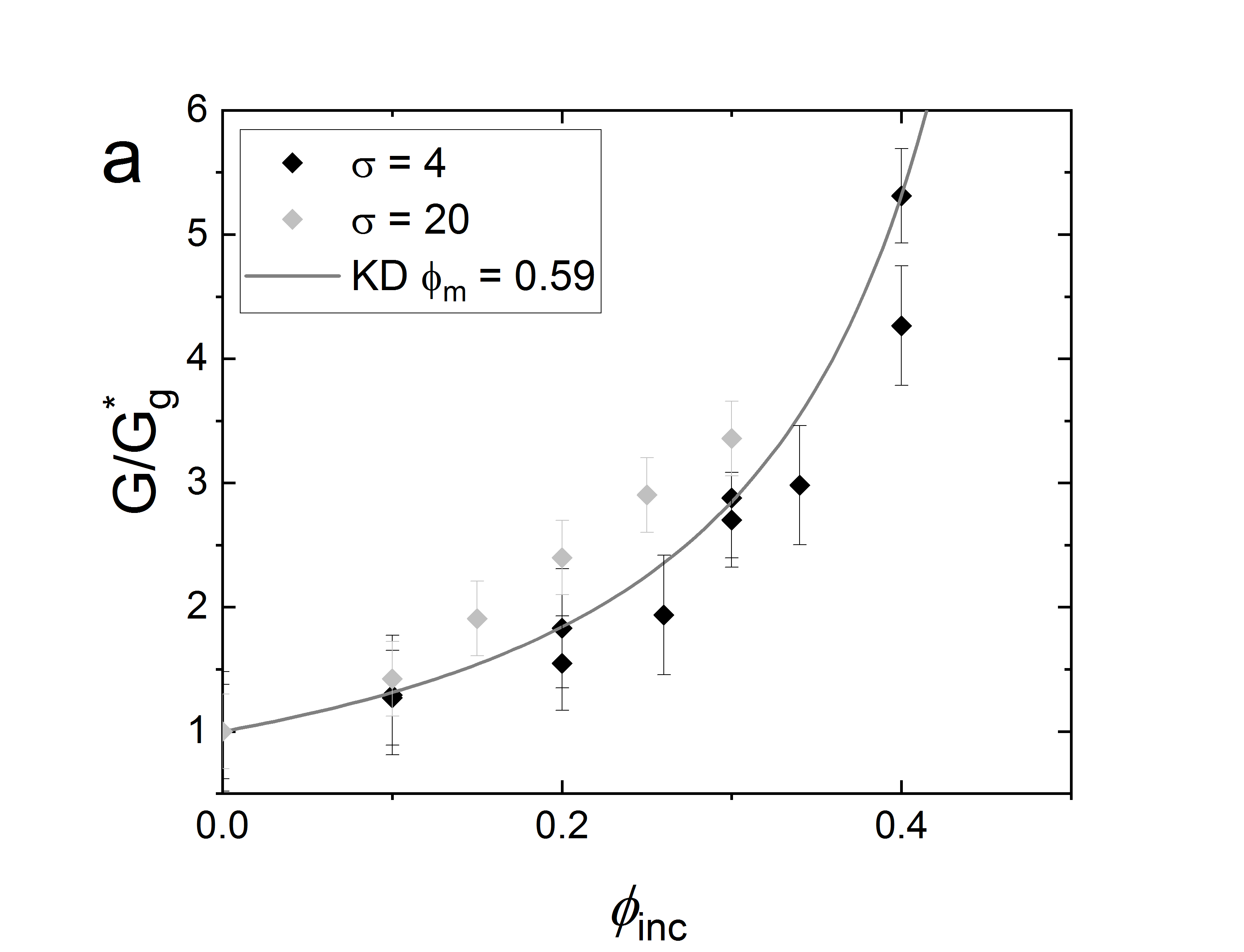}
\includegraphics[width=0.45\textwidth]{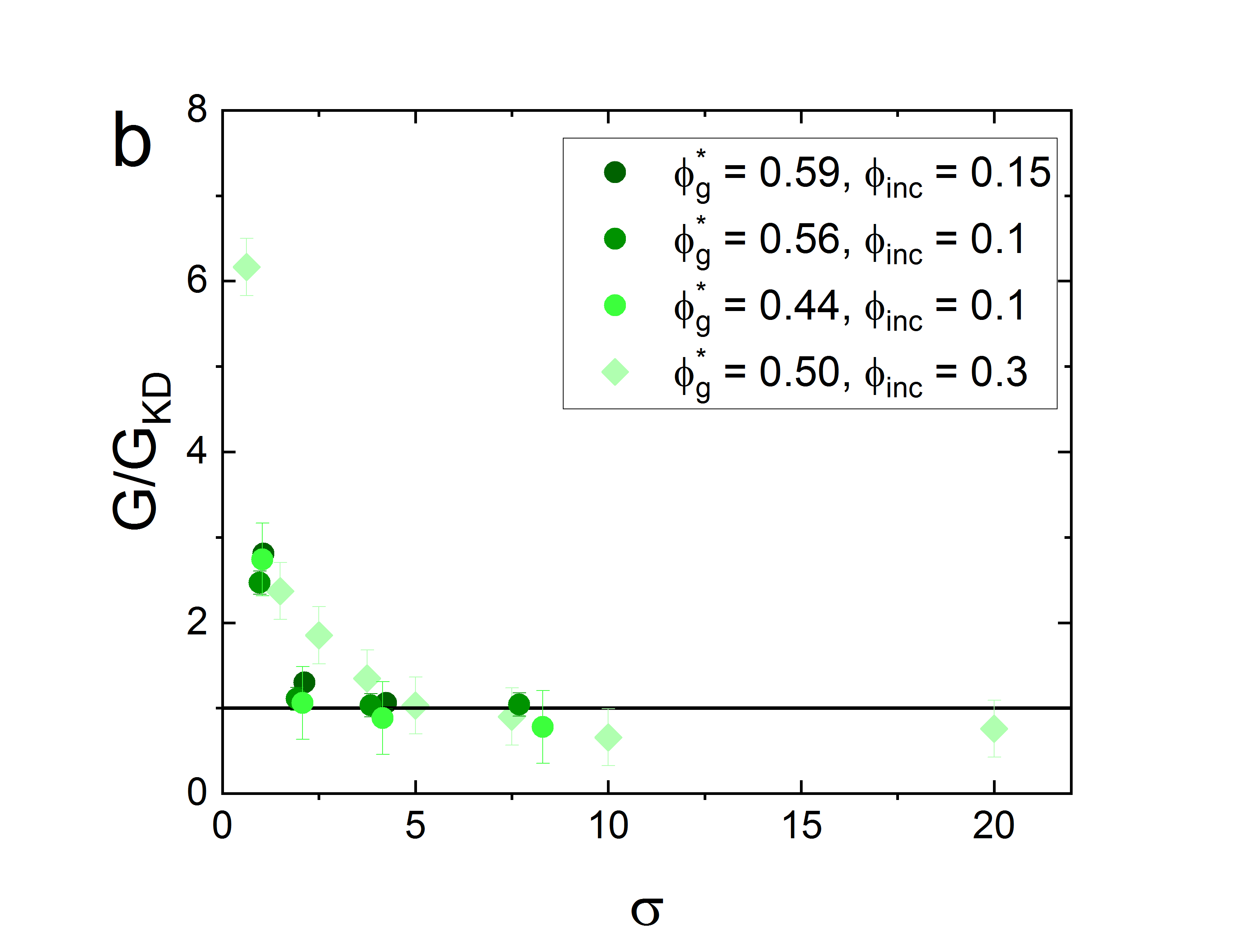}
\includegraphics[width=0.45\textwidth]{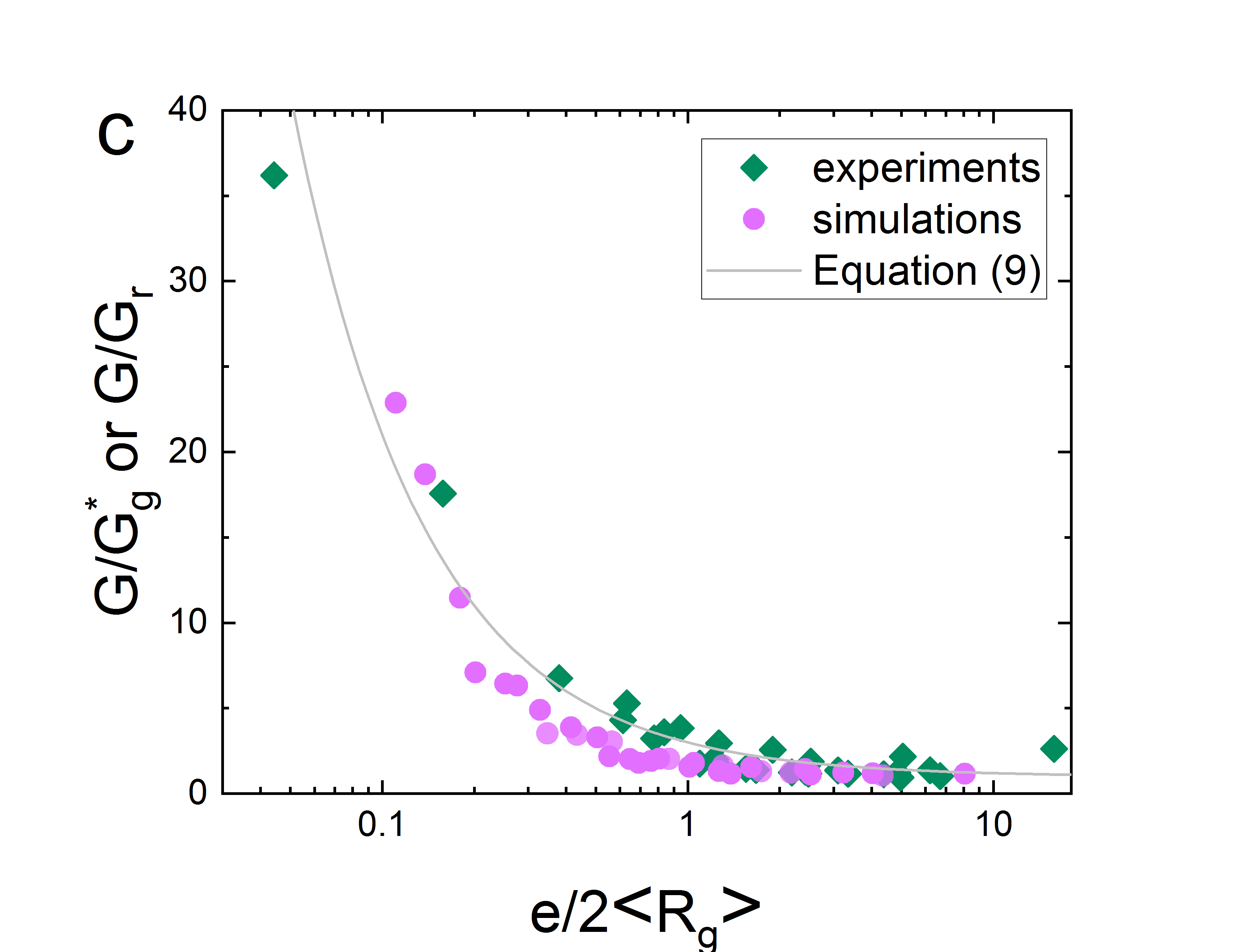}

\caption{a)  Elastic modulus of loaded emulsion divided by the elastic modulus of the embedding emulsion $\phi_{g}^*=0.5$ as a function of the volume fraction of inclusions $\phi_{inc}$, for two size ratios: $\sigma$ = 4 and $\sigma$ = 20. The solid line shows the shear elastic modulus $G_{KD}$ according to the Krieger Dougherty law with $\phi_{m}$ = 0.59 and $x$=2.5. b) Plot of $G/G_{KD}$ for different loaded emulsions (characterized by $\phi_{g}^*$ and $\phi_{inc}$) as a function of the size ratio. c) Elastic modulus normalised by the inclusion free modulus $G/G_g^*$ for experiments and $G/G_r$ for simulations as a function of $e/2\langle R_g \rangle$. The continuous line is given by Equation \ref{strengthening}. }
\label{fig:experimentalRheo}
\end{figure}

We measure the elastic modulus of the composite gel at different $\phi_{inc}$ for inclusion sizes corresponding to $\sigma = 4$ and $\sigma = 20$ as calculated with the average emulsion drop size. As it could be expected, the elastic modulus of the composite gel increases with increased $\phi_{inc}$. In the experiments we expect that the boundary condition is a no-slip condition \cite{Mahaut2008, Vu_2010}, and we compare to the KD model with $x = \frac{5}{2}$. The two  measurements of $G/G_g^*$ are shown in Fig.~\ref{fig:experimentalRheo}a as a function of $\phi_{inc}$. The red line shows the KD prediction with $\phi_m = 0.59$ that describes the results well.  

Having seen that for $\sigma = 4$ the KD model works well, we explore the effect of varying $\sigma$. In Fig.~\ref{fig:experimentalRheo}b, the experimental composite gel elastic moduli are normalised using the KD prediction as a function of inclusion size. For $\sigma$ larger than $2$ or $3$, the enhancement of the inclusions to the elastic modulus compares well with the KD model. However, at smaller $\sigma$ we observe an upturn. As the inclusion size approaches the average drop size we see a strong increase in the elastic modulus. Such a strong strengthening behaviour revealed for $\sigma \simeq 1$ was not observed by Gorlier et al. \cite{Gorlier_SoftMatter_2017} for particle loaded foams. This suggests that the nature of interactions between the particles, i.e. here attractive vs repulsive for liquid foams, is a crucial parameter for strengthening. Therefore, and in a more general way, it appears that tuning $\sigma$ in colloidal gels can be of great interest to optimize their strength.

The $\sigma$ effect highlighted in Fig.~\ref{fig:experimentalRheo}b is reminiscent of the simulation results shown in Fig.~\ref{fig:takingoutinclusions}c. In order to compare more closely with our numerical results, we plot in Fig.~\ref{fig:experimentalRheo}c $G/G_g^*$ and $G/G_r$ as a function of the gap size between inclusion particles, i.e. $e/2\langle R_g \rangle = \sigma((\frac{\phi_m}{\phi_{inc}})^{\frac{1}{3}}-1)$.
As can be seen, the agreement with numerical results is remarkable, which gives significant insight into the finite size effects highlighted with the attractive emulsion: small inclusions (i.e. $e/2\langle R_g \rangle < 1$) are allowed to be embedded within the gel particle backbone and to provide optimal strengthening by supporting the backbone, while larger inclusions (i.e. $e/2\langle R_g \rangle \gg 1$) are not included into the backbone and behave like classical solid inclusions, providing strengthening in agreement with continuum mechanics. For practical purposes, the measured finite size effect can be reasonably described by the following approximate function:

\begin{equation}
    \frac{G}{G_g^*}  \simeq 1 + \frac{2}{\sigma} \left ( \left( \frac{\phi_m}{\phi_{inc}}\right)^{1/3} -1 \right)^{-1}
\label{strengthening}
\end{equation}

Although this empirical equation can be useful to evaluate the magnitude of finite size effects for $e/2\langle R_g \rangle < 2-3$, we stress that (1) it concerns attractive gels only, and (2) it does not capture properly strengthening in the limit $e/2\langle R_g \rangle \gg 1$, which is better described by the KD equation.

\section{Conclusion}
\label{sec4}

We have combined rheology computer simulations with experiments to explore the elastic modulus of composite attractive gels. Our work reveals significant finite size effects in the mechanical response of such systems and gives physical insight into their origin. These effects depend on the total volume fraction ($\phi_{\text T}$) in the system (gel + inclusions): At low total volume fraction ($\phi_T < 0.6$) the addition of inclusions leads to a weak stiffening of the material, below the predictions of continuum mechanics (KD model). However at $\phi_T = 0.6$ the effect of the inclusions becomes strongly size dependent. 

With large inclusions $\sigma \gg 1$ the elastic modulus of the attractive composite gels is correctly described using the KD prediction. On the contrary, as the inclusion size decreases we observe considerable stiffening of the material. These findings can be explained by the effects of the inclusions, they can change the structure of the gel backbone and the average interactions, as the gel-inclusion interactions are weaker than the gel-gel interactions. If the inclusions are large the gel backbone structure is only weakly modified by the presence of the inclusions, however once they become small they participate actively in the gel structure and hence the mechanical properties. 

The degree of stiffening is well accounted for by the distance between the inclusions. Once the distance between the inclusions becomes comparable to the size of the gel particles finite size effects dominate and continuum mechanics approaches do not capture the mechanical properties of the material. This result is confirmed by both the simulations and the experiments. This despite several differences between systems studied in the simulation and experiments, in particular in terms of gel particle polydispersity, interactions and effective modulus of the particles.
This suggests that the reported findings concern a broad range of composite materials. It should be stressed however that the attractive nature of interactions in the studied gels seems to be necessary for those finite size effects to be significant.

The results provide guidance for the control over the mechanical properties of colloidal gels using inclusions. However they can also help understand the many practical materials where solid inclusions are found in emulsions or gels, and the size ratios of the particles is no longer large. Such composite materials are present in drilling fluids, food or cosmetic products.

\begin{acknowledgments}
We thank Stefan Egelhaaf and Véronique Trappe for insightful discussions. We thank Sandrine Mariot for help with the use of the confocal microscope. The funding for this project was provided by ESA (Soft Matter Dynamics) and CNES (Hydrodynamics of wet foams). CFC and GF thank the LabEx PALM for support (Project  MuSa).  
\end{acknowledgments}

\bibliographystyle{naturemag}
\bibliography{Inclusion}

\end{document}